\documentclass{article}
\usepackage[authoryear]{natbib}
\usepackage{arxiv}
\usepackage[T1]{fontenc}
\usepackage{lineno}
\usepackage{fix-cm}
\usepackage{graphicx}
\usepackage{multirow}
\usepackage{amsmath,amssymb,amsfonts}
\usepackage{amsthm}
\usepackage{array}
\usepackage[title]{appendix}
\usepackage{natbib}
\usepackage{xcolor}
\usepackage{booktabs}
\usepackage{listings}
\usepackage{hyperref}
\usepackage{arydshln}
\usepackage{fancyvrb}
\usepackage[mathscr]{eucal}
\usepackage[figuresright]{rotating}
\usepackage{textcomp}%
\usepackage{manyfoot}%
\usepackage{algorithm}%
\usepackage{algorithmicx}%
\usepackage{algpseudocode}%
\usepackage{setspace} 
\usepackage{lineno}
\usepackage{hyperref}
\usepackage{cleveref}
\hypersetup{
    colorlinks=true,
    citecolor=blue!80!black,    
    linkcolor=blue!80!black,    
    urlcolor=blue!80!black      
}
\usepackage{booktabs}
\usepackage{rotating}
\usepackage{caption}
\usepackage{tikz}
\usepackage{makecell} 
\usepackage{appendix}
\usepackage{url}

\newcommand{\hypothesis}[2]{
    \par\vspace{0.5em}
    \noindent\textbf{H#1:}\;\; \textit{#2}
    \par\vspace{0.5em}
}

\usepackage{indentfirst}
\title{Dissipation of Debt Financing Privilege on Corporate AI Washing: Evidence from China}

\author{
  \textbf{Congluo Xu}\thanks{Contact Email \texttt{xucongluo@stu.scu.edu.cn}}\\
  Business School\\ Sichuan University\\
  Chengdu, China, 610065
  \\[2ex]
  \textbf{Xiangsheng Zheng}\\
 School of Accounting\\ SWUFE\\
  Chengdu, China, 611130
  \And
  \textbf{Jiuyue Liu} \\
  School of Economics\\ Sichuan University\\
  Chengdu, China, 610065
  \\[2ex]
  \textbf{Ziyang Li}\\
  Business School\\ Sichuan University\\
  Chengdu, China, 610065
}

\begin{document}

\maketitle

\begin{abstract}
The rapid development of artificial intelligence motivates firms to engage in AI washing. This study examines whether strategic policy shocks increase debt financing costs for such firms. Leveraging China's 14th Five Year Plan as a quasi natural experiment, we identify AI washing through the residual between AI narrative intensity and patent output. External validation confirms this decoupling reflects strategic deception evidenced by subsidy extraction and future regulatory violations rather than benign ambition, supporting its validity as an AI washing proxy. Difference in differences estimations reveal that AI washing firms experience a 12.5 basis point relative increase in debt financing cost afterward. Joint estimation confirms simultaneous adjustments across financing and innovation margins. Management shareholding and analyst attention amplify the penalty while supply chain concentration and bank proximity attenuate it. Results remain robust across checks. Our findings illuminate how macro level policy shocks activate market discipline in emerging market debt markets.
\end{abstract}

\keywords{Debt Financing Cost, Artificial Intelligence Washing (AI Washing), Quasi-natural Experiment, National Strategic Planning, China's 14th Five-Year Plan}

\section{Introduction}

Technological innovation and capital market efficiency are typically regarded as mutually reinforcing, yet the inherent opacity of emerging technologies generates significant information asymmetry that impedes efficient resource allocation \citep{brown2019does}. The rapid advancement of artificial intelligence (AI) represents one of the most profound technological shifts of this century, offering considerable economic opportunities for firms. However, this considerable potential has also given rise to corporate AI washing, defined as the systematic decoupling between AI-related narratives in corporate disclosures and substantive AI outcomes \citep{song2026ai}.

This phenomenon originates not within the domain of technological innovation, but within corporate ethics and environmental reporting. As highly mature capital markets in Europe and America prioritize elevated societal objectives such as governance ethics and environmental sustainability \citep{doh2015csr, kolk2016social, kumar2021cross}, these considerations transition from peripheral concerns to central priorities. Consequently, firms opportunistically engage in deceptive practices such as CSR washing \citep{pope2016csr}, greenwashing \citep{marquis2016scrutiny}, and reputation washing \citep{rozeboom2025corporate}, exploiting complexity to mislead stakeholders \citep{crilly2012faking} and earn profits. Yet, the necessity to investigate AI washing stems not only from its novelty as a variant of such washing phenomenon, but from its distinctive penetration into corporate operational core.

Innovation capability constitutes the cornerstone of firm viability, and its decrease will undermine long-term operational sustainability \citep{zahra1993environment, scherer2020corporate, li2022can}. Unlike previous ethical washing, which threatens reputation, AI washing manifests as strategic decoupling between innovation narratives and actual technological output, directly eroding the fundamental capabilities underpinning competitive advantage. In emerging economies such as China, where only a subset of mature firms possess the requisite capacity to pursue genuine elevated objectives, innovation constitutes a foundation and priority to them all. AI washing not only compromises individual corporate competencies, but also distorts resource allocation across the broader market ecosystem, thereby impeding economy-wide sustainable development, which should be taken on heightened significance.

The literature on AI washing remains empirically fragmented, with existing studies predominantly focusing on implications for technological development per se or interaction mechanisms with ESG (Environmental, Social, and Governance). \citet{sun2026unveiling} identifies an inverted U-shaped relationship between AI washing and the technological gap among firms. \citet{xu2026illusion} demonstrates that AI washing significantly inhibits green technology innovation by diverting managerial attention from environmental concerns. \citet{zhou2026contractual} establishes that ESG executive compensation incentives constitute a key organizational mechanism for curbing AI washing. These studies illuminate how AI washing undermines technological progress and how firms attempt to mitigate it through organizational design. Yet by conceptualizing AI washing exclusively as an intra-organizational phenomenon with technological or governance implications, these studies preclude examination of corporate financing implications and the potential for policy-induced behavior discipline.

This omission is theoretically significant because several studies have noticed that national policy shocks can fundamentally alter the information environment in capital markets, redirecting market and public attention from other domains toward policy-focused sectors and inducing capital reallocation across them \citep{keister2004capital, cong2019credit, wang2022public, liu2022capital}. This study therefore examines the capital allocation consequences of such policy interventions in the context of AI washing, with particular attention to corporate debt markets where AI washing exerts pronounced distortions on capital allocation efficiency. Such attention is because of creditors' fixed-payoff structure and prohibitively high screening costs regarding complex technological capabilities, and they necessitate heavy reliance on corporate disclosures \citep{bester1985screening, demerjian2024positive}, rendering debt financing particularly susceptible to narrative manipulation and sustaining information asymmetry wherein genuine innovators and opportunistic impostors remain observationally indistinguishable \citep{takalo2010adverse}. Consistent with this theoretical prediction, \citet{liu2026impact} has demonstrated that AI washing significantly facilitates access to bank loans, corroborating systematic credit misallocation. But whether strategic policy can disrupt this misallocation and enable debt market participants to impose differential financing costs upon AI washing firms, still remains an open question of considerable theoretical and practical import.

China's 14th Five-Year Plan, released in 2021, provides a quasi-natural experiment environment that directs unprecedented market attention toward AI \citep{stern2023china}, in which ``New Generation of AI'' is ranked first among the seven key directions in the forefront of science and technology. This policy shock enables an investigation into whether such interventions induce differential pricing in debt markets, wherein firms engaging in AI washing experience significantly greater increases in debt financing cost relative to genuine AI adopters, thereby narrowing the pre-policy financing gap.

To empirically examine this question, this study employs a Difference-in-Differences (DID) design exploiting the release of the 14th Five-Year Plan. We operationalize AI decoupling as the systematic residual of AI-related narrative intensity relative to actual AI patent output, controlling for firm-level variation and industry fixed effects. We then validate its external fraudulence and employ it as a proxy for AI washing. We also construct a comprehensive dataset encompassing A-share listed firms from 2015 to 2024. The empirical sample comprises firms listed prior to the policy shock with observable AI disclosure histories preceding the intervention, enabling clear delineation of treatment and control groups.

We document a significant positive association between corporate AI washing and relative increases in debt financing cost following the policy shock. While descriptive statistics indicate that these firms maintain lower absolute levels of debt financing cost post-shock, the inter-group difference diminishes substantially and even loses statistical significance. Baseline estimations reveal that firms habitually engaged in AI washing pre-policy experience significantly greater increases in debt financing cost relative to the control group, with the differential increment amounting to 12.5 to 19.3 basis points. This magnitude, equivalent to approximately 8\% to 11\% of the sample mean of 1.69\% and representing 0.08 to 0.12 standard deviations, constitutes a cost increase comparable to that from regulatory penalties \citep{gong2021punishment}, thereby reflecting the dissipation of their pre-policy financing privilege. These findings withstand rigorous robustness checks addressing measurement bias, selection bias, and competing policy confounds. Seemingly unrelated regressions (SUR) estimation further confirms joint significance across debt financing cost, credit availability, AI narrative intensity, and patenting outcomes, revealing the multi-dimensional nature of market discipline. The constraining effect is disproportionately concentrated among firms with limited non-self patent citations and weak disruptive innovation capacity, whereas informal institutions, specifically commercial guild affiliations and government connections, function as buffering governance mechanisms. Further analyses of moderating mechanisms reveal that the punitive effect is amplified among firms characterized by concentrated management shareholding or heightened external analyst attention, yet attenuated among those embedded in dense supply-chain networks or exhibiting high geographical proximity to banks. These findings underscore the dual emphasis on deep capability signals and relational ties within China’s socio-commercial institutional environment.

We make four primary contributions to the literature.

First, we contribute to the measurement of corporate AI washing by extending the residual-based approach from the earnings management literature to AI decoupling. While prior studies employ this method to capture discretionary accruals or real earnings manipulation, we adapt it to quantify the systematic divergence between AI-related narrative intensity and substantive patent output. External validation demonstrates that this decoupling is strategically consequential rather than random measurement error. Firms with higher residuals exhibit a greater likelihood of receiving government innovation subsidies and a higher probability of future regulatory violations, establishing that symbolic disclosure facilitates resource extraction and conceals genuine misconduct. The temporal pattern of regulatory responses further reveals that AI decoupling evades near-term detection yet persists until cumulative evidence prompts subsequent investigation. These patterns validate the residual as an effective proxy for AI washing and provide a replicable empirical foundation for research on corporate technological disclosure integrity.

Second, we advance the policy shock literature by establishing that strategic priority signals can function as information-coordination devices even in the absence of explicit mandates or punitive enforcement. Our evidence demonstrates that a macro-level shock generates heterogeneous micro-level debt pricing outcomes contingent on pre-shock AI decoupling behavior, because the shock increases the marginal value of AI verification and induces creditors to concentrate screening resources on distinguishing substantive patent output from symbolic claims. This finding reveals how policy realigns creditor screening incentives with policy-focal domains through market attention rather than through direct credit allocation or regulatory coercion.

Furthermore, we isolate the operative transmission channel and establish that the documented treatment effect operates through information-driven market discipline rather than contemporaneous regulatory enforcement. Our moderating analysis reveals that the interactions between AI washing and regulatory violations, as well as between AI washing and regulatory inquiries, are statistically indistinguishable from zero. While regulatory actions independently increase financing costs, they do not differentially amplify the penalty for AI washing firms. This null result rules out direct regulatory punishment as the driving force. Instead, the dissipation of financing privilege reflects endogenous market discipline activated through information environment restructuring, wherein creditors autonomously reallocate attention toward AI capability verification rather than responding to concurrent punitive measures. This information channel constitutes the primary mechanism through which macro-level strategic shocks activate micro-level market discipline.

Finally, we challenge the conventional view that emerging-market debt markets lack the capacity to assess complex technological claims \citep{matolcsy2008association, wei2022r}. Our evidence reveals a dual institutional complementarity distinctive to China. Policy-induced attention shocks activate latent creditor screening capabilities, yet indigenous relational governance structures simultaneously attenuate market discipline through informal buffers that substitute for direct technological verification. This coexistence of hard-information scrutiny and relational buffering demonstrates that emerging-market capital allocation efficiency depends not on replicating Western-style arm's-length contracting but on recognizing how policy shocks and network-based contracting mechanisms jointly shape creditor behavior.

The remainder of this paper is organized as follows. Section \ref{Literature Review} situates our study within the literature on AI washing and attention reallocation, developing the testable hypotheses. Section \ref{Methodology} describes the data and identification strategy. Sections \ref{Empirical Results}$\sim$\ref{Further Explorations} present the external validation, empirical results and further analyses. Section \ref{Conclusion} discusses and concludes.

\section{Literature Review and Hypothesis Development}\label{Literature Review}

\subsection{AI Washing}\label{2.1}

Neo-institutional theory \citep{meyer1977institutionalized} reveals systematic decoupling between formal organizational structures and actual operational practices. When external stakeholders evaluate firms based on specific performance signals, departments accountable for those signals exhibit more pronounced decoupling \citep{behnam2011accountability, crilly2012faking, shi2018regulatory}. AI washing represents such decoupling in the context of artificial intelligence \citep{bernini2025measuring}. Prior research draws direct parallels to greenwashing, documenting institutional isomorphism and comparable adverse consequences between these two forms of symbolic misconduct \citep{seele2022greenwashing}. Institutional pressures for environmental legitimacy and employee welfare drive greenwashing \citep{testa2018internalization}, whereas AI washing emerges from capital market pressures for artificial intelligence capabilities \citep{xing2026ai}. Both practices generate extra costs through capital misallocation, as stakeholders cannot distinguish symbolic claims from substantive performance.

AI washing stems from the inherent unverifiability of AI technology and external stakeholders' limited capacity to assess technical substance. Corporate innovation exhibits substantial heterogeneity in trajectories and timing \citep{garcia2005uses, katila2008effects}, precluding universal benchmarks for optimal innovation \citep{payne2006examining}. Yet market participants evaluate firms based on contemporaneous performance signals \citep{busenbark2017foreshadowing}, rewarding those demonstrating superior observable outcomes. Simultaneously, operational path dependence discourages substantive AI investment, as large-scale technological transformation disrupts existing operational routines and accumulated capabilities \citep{argyres2004r, argyres2020organizational}. These conditions collectively reduce the opportunity costs of decoupled disclosure, enabling firms to exploit information asymmetry by claiming unobservable technical capabilities without bearing commensurate implementation costs. Stakeholders possess limited technical literacy regarding AI applications \citep{sartori2023minding}, relying primarily on observable symbolic claims to infer organizational competence \citep{lajoie2025content}. The absence of verifiable capability signals implies that AI washing often imposes limited reputational or legal costs on disclosing firms \citep{xing2026ai}, and may even yield positive legitimacy gains through enhanced resource access. Regulatory frameworks governing AI disclosure remain nascent, and the legal consequences of symbolic misrepresentation are presently indeterminate \citep{schmitt2022mapping, carey2026regulating}. These institutional voids further attenuate the expected costs of decoupling.

Although AI washing is conceptually established, the measurement literature remains underdeveloped, with extant studies relying primarily on case studies or small-sample content analysis. Recent large-sample evidence nevertheless documents that AI washing confers significant debt financing privileges, with \citet{liu2026impact} demonstrating that symbolic disclosure substantially facilitates bank loan access, corroborating systematic credit misallocation toward decoupled firms. However, precise quantification of these economic consequences remains scarce. More critically, whether this debt financing privilege persists unconditionally or dissipates under exogenous information shocks remains unexamined in the existing literature.

\subsection{Policy Shock Shapes Attention Reallocation}\label{2.2}

While previous work documents that AI washing attracts credit inflows and gains relative debt financing privilege \citep{liu2026impact}, this evidence derives primarily from banking relationships, leaving the broader implications for aggregate corporate debt financing cost underexplored. Limited attention theory posits that investors face binding cognitive constraints in informationally rich markets \citep{hirshleifer2003limited}, necessitating selective focus on observable attributes rather than substantive due diligence \citep{loewenstein2025economics, mackowiak2023rational}. These constraints bind creditor attention to specific performance signals, rendering corporate debt financing cost sensitive to variation in the information environment.

Exogenous shocks cause a variation and operate through distinct channels. Risk-elevating events, such as disasters, expand attentional depth while preserving breadth as investors process higher stakes \citep{huynh2023panic}. Administrative reorganizations and regulatory reforms, by contrast, operate through mandatory institutional changes that compel attention reallocation \citep{bertelli2015mass, abad2019informational}. These interventions systematically redistribute attention by altering the information content of ratings or intensifying media coverage of focal domains, thereby shifting creditors' cost-benefit calculus of screening and monitoring \citep{federico2025trade, cong2019credit}. However, whether macro-level strategic policy absent mandatory restructuring generates comparable attention reallocation in corporate debt markets remains unclear.

China's 14th Five-Year Plan differs fundamentally from these precedents along three dimensions. It has not triggered comparable administrative restructuring \citep{wirtz2020dark}; it does not constitute a narrow industrial policy \citep{dunleavy2025data}; and it lacks a fully articulated regulatory framework with punitive enforcement \citep{af2023discursive}. The Plan therefore represents a pure signal of national strategic priority that alters the information environment without coercive institutional mechanisms. We posit that such strategic priority signals operate through incentive-driven attention reallocation. By elevating artificial intelligence to strategic national priority, the Plan increases the marginal value of AI capability information for creditors. Consequently, creditors voluntarily intensify screening efforts toward AI-related signals to exploit informational advantages in a policy-focal domain. Whether this incentive-driven mechanism disciplines corporate debt financing cost for firms with decoupled AI disclosures remains an open empirical question that this study addresses.

\subsection{Hypothesis Development}\label{2.3}

Information asymmetry between firms and creditors distorts debt capital allocation \citep{stiglitz1981credit, healy2001information}, prompting creditors to rely on observable signals to assess corporate quality and price debt contracts. Innovation capability serves as a primary signaling device through which firms convey quality to external capital providers \citep{moss2015effect}, determining creditor assessments of default risk and associated risk premiums \citep{mancusi2014r, cerqueiro2017debtor}. Annual reports function as the primary channel for transmitting such signals, summarizing innovative activities following the fiscal year \citep{krishnan2009recent} while complementing the limited information regarding innovation investments in standardized financial statements \citep{merkley2014narrative}. When AI-related narrative disclosures deviate from substantive patenting outcomes, the resulting decoupled signals obscure true technological capabilities. Creditors' limited attention impedes verification of these claims against patent records \citep{hottenrott2016patents}, prompting reliance on readily available textual signals. This reliance on surface-level indicators persists until national policy elevates AI to strategic priority, fundamentally altering the information environment.

China's 14th Five-Year Plan represents a discontinuous regime shift rather than gradual refinement, constituting an unprecedented elevation of AI to strategic national priority \citep{roberts2021chinese}. \autoref{Policy Comparison} indicates that the policy emphasis on artificial intelligence has intensified significantly rather than representing incremental iteration.

\begin{table}[htbp]
  \centering
  \caption{\centering Comparison of Artificial Intelligence Significance Across Five-Year Plans}
  \label{Policy Comparison}
  \vspace{0.1em}
  \renewcommand{\arraystretch}{1.15}
  \begin{tabular}{l*2{>{\arraybackslash}p{6cm}}}
    \toprule
    \textbf{Dimension} & 
    \makecell[c]{\textbf{13th Five-Year Plan}\\\textbf{(2016--2020)}} & 
    \makecell[c]{\textbf{14th Five-Year Plan}\\\textbf{(2021--2025)}} \\
    \midrule
    \textbf{Strategic Position} & 
    One of ``Strategic Emerging Industries'' in the Information Technology Sector & 
    Elevated to a \textbf{National Strategic Priority} With Dedicated Policy Focus \\
    \addlinespace
    \textbf{Expression Frequency} & 
    Sporadic References to ``Intelligence'' Without Specific AI Emphasis & 
    \textbf{Dense and Explicit Mentions} of ``AI'' Throughout Strategic Chapters \\
    \addlinespace
    \textbf{Technical Objectives} & 
    Preliminary Technology Tracking and Basic Ecosystem Establishment & 
    Designated as \textbf{National Megaprojects} With Breakthrough Mandates \\
    \addlinespace
    \textbf{Application Scope} & 
    Localized Pilot Programs Restricted to Demonstration Zones & 
    \textbf{Comprehensive Industrial Coverage} and Cross-Scenario Deep Integration \\
    \addlinespace
    \midrule
    \textbf{Literature Support} &
    \citet{hong2017reading} &
    \citet{poo2021innovation}; \citet{khanal2025development} \\
    \bottomrule
    \addlinespace[0.1em]
    \multicolumn{3}{l}{\footnotesize Note: China's 13th Five-Year Plan marked the inaugural policy references to artificial intelligence-related mentions.} \\
  \end{tabular}
\end{table}

We posit that this national strategic initiative induces an attention crowding-in mechanism. By elevating artificial intelligence to strategic national priority, China's 14th Five-Year Plan reallocates limited creditor attention toward AI-related signals. This concentration of cognitive resources on the policy-focal domain permits creditors to enhance screening depth. Consequently, firms with historically decoupled AI disclosures face higher detection probability as creditors allocate increased attention to verifying substantive AI capabilities against patent records. Creditors respond by imposing adverse debt pricing through elevated risk premia, thereby increasing debt financing cost for these firms relative to genuine AI adopters and eroding their pre-policy financing advantage. Accordingly, we hypothesize that:

\hypothesis{1}{China's 14th Five-Year Plan relatively increases the debt financing cost of pre-policy AI washing firms.}

The identification risk and relative cost increase documented in \textbf{H1} alter not merely debt pricing but also corporate investment incentives and external financing quantities. Rational firms facing heightened detection probability optimally converge toward substantive capability demonstration. Specifically, firms previously reliant on AI washing have strong incentives to reduce symbolic AI disclosure intensity to avoid detection, while simultaneously increasing substantive AI patenting to rebuild technological credibility with capital providers.

However, creditor responses to recognized information risk extend beyond risk premium adjustments to the quantity dimension of debt contracting. Creditors respond to identified decoupling by imposing stricter covenant terms \citep{bharath2008accounting} or reducing credit line commitments \citep{balakrishnan2019bank}. Because information asymmetry prevents market clearing through interest rate adjustments alone, the equilibrium exhibits credit rationing alongside elevated risk premiums \citep{stiglitz1981credit}. This financing constraint intensifies resource scarcity, compelling firms to reallocate limited capital toward verifiable technological investment and away from symbolic disclosure activities. Consequently, pre-policy AI washing firms experience a dual adjustment: constrained narrative strategies and expanded patent portfolios, coupled with both higher marginal financing costs relative to comparable firms and constrained debt financing flows. Accordingly, we hypothesize that:

\hypothesis{2}{Pre-policy AI washing firms exhibit reduced AI narrative intensity and increased AI patenting following the policy implementation, simultaneously facing higher relative financing costs and constrained debt financing flows.}

The magnitude of relative debt financing cost adjustments documented in \textbf{H1} varies with the information environment in which creditor screening occurs. Because creditor attention is not directly observable \citep{karlan2009observing}, we proxy its intensity using management shareholding \citep{hong2021lender} and analyst attention level \citep{cheng2008analyst}. Management shareholding captures the internal governance dimension of screening effort, as creditors scrutinize insider equity positions to assess the credibility of managerial disclosure. analyst attention level captures the external information-network dimension, as analyst monitoring directs and amplifies market attention toward specific firms. The attention crowding-in mechanism operates through these two complementary channels.

First, internal governance signals shape the credibility of managerial disclosure when screening depth increases. Management shareholding typically alleviates debt financing constraints by aligning managerial incentives with firms value maximization \citep{JENSEN1976305}. However, when AI washing coexists with substantial management shareholding stakes, creditors interpret this combination as informed strategic misconduct rather than uninformed disclosure error. Managers with significant equity positions possess both the technical sophistication to understand AI capabilities and the strategic incentive to exploit information asymmetry. As creditor attention concentrates on AI-related signals, these governance characteristics become more salient in risk assessments. The informed insider status transforms symbolic misrepresentation from potential negligence into credible evidence of intentional narrative manipulation, prompting creditors to impose steeper relative risk premia.

Second, analyst attention level amplifies the reputational consequences of AI washing and coordinates creditor scrutiny across the information network. When analysts concentrate attention on specific firms, this concentration functions as a signal of interest that directs creditor attention toward high-visibility firms \citep{cheng2008analyst}. Elevated analyst attention level increases the reputational costs of screening failures, drawing creditors to redirect attention toward these firms. Consequently, AI washing firms operating under intense analyst attention level experience more pronounced relative financing cost penalties as the attention crowding-in effect amplifies. Accordingly, we hypothesize that:

\hypothesis{3}{The positive association between pre-policy AI washing and post-policy relative debt financing cost is more pronounced for firms with higher management shareholding and greater analyst attention level.}

The attention crowding-in mechanism documented in \textbf{H3} does not operate uniformly across all firms. Its punitive effect depends on the availability of alternative assurance mechanisms that substitute for direct technological verification.

China's institutional environment features relational structures that function as alternative assurance devices when formal verification channels prove imperfect \citep{xin1996guanxi, zhou2003embeddedness}. Supply chain concentration creates dense, repeated interactions between specific buyers and suppliers, generating relational capital that serves as implicit collateral against negative signals. When creditors assess firms with concentrated supply chains, they infer the existence of verified, ongoing commercial relationships that constrain symbolic misconduct. The accumulated trust between transaction partners buffers the relative debt financing cost increase by substituting for direct AI capability screening.

Geographic proximity to banking institutions generates a complementary buffering mechanism. Proximate bank presence facilitates soft information production through repeated interactions and community-based monitoring \citep{agarwal2010distance, del2020soft}, providing alternative assurance channels. When firms maintain close geographic ties to dense banking networks, creditors access informal intelligence regarding firm capabilities through relationship lending channels, attenuating the necessity of penalizing decoupling through risk premium adjustments. This geographic dimension of banking support operates as a spatially embedded governance mechanism that attenuates the attention crowding-in effect. Accordingly, we hypothesize that:

\hypothesis{4}{The positive association between pre-policy AI washing and post-policy relative debt financing cost is attenuated for firms with higher supply chain concentration and greater geographic proximity to banking institutions.}

\section{Methodology}\label{Methodology}

\subsection{Sample and Data}

This study constructs the sample using Chinese A-share listed firms from 2015 to 2024 to align with the time frame for the 14th Five-Year Plan implementation. To examine the above hypotheses, we integrate the financial characteristics, annual reports and patent data of Chinese A-share listed firms. Financial statement data, corporate governance characteristics including board structure and list age, and textual AI disclosure measures derive from the China Stock Market and Accounting Research Database (CSMAR) and annual reports. AI patent data comes from the Artificial Intelligence Patent Research Database (AIPD) on the Chinese Research Data Services Platform (CNRDS).

To ensure sample quality and mitigate confounding influences, we implement the following screening procedures: (1) dropping observations from the financial sector given their distinct regulatory capital requirements and contracting environments; (2) dropping observations receiving ST, ST*, or PT,\footnote{\;ST (Special Treatment), ST* (Special Treatment with imminent delisting risk), and PT (Particular Transfer) are regulatory designations assigned by the China Securities Regulatory Commission to firms experiencing financial distress, abnormal trading volatility, or failure to meet minimum listing requirements.} as these firms face unique financial distress conditions and mandatory disclosure regimes that may confound inferences regarding voluntary AI disclosure behavior; (3) dropping observations with missing values for key regression variables including AI washing measures, debt financing cost, and corporate-level control variables; (4) applying industry-year mean imputation to observations with non-critical missing values; (5) winsorizing the top and bottom 1\% of all the continuous variables. The final sample comprises 34,924 firm-year observations.

\subsection{Variable Definitions}

\subsubsection{Debt Financing Cost Measure}

Following \citet{zou2008debt} and \citet{wang2022government}, we measure debt financing cost (Debt FC) as the ratio of total debt-related expenditures to end-of-period total liabilities. Firms employ heterogeneous debt instruments including bank loans, corporate bonds and commercial paper. These instruments impose costs beyond contracted interest rates such as commitment fees, origination charges and administrative expenses. Given that interest expense alone excludes these non-interest cost components, we employ the comprehensive ratio \autoref{eqDebtFC} to capture the all-in cost of corporate debt financing.
\begin{align}\label{eqDebtFC}
    \mathrm{Debt\;FC} = \frac{\mathrm{Interest\;Expense} + \mathrm{Service\;Fees} + \mathrm{Other\;Financial\;Expenses}}{\mathrm{End\;of\;Period\;Total\;Liabilities}}
\end{align}

\subsubsection{AI Washing Treatment Status Measure}

Following \citet{athey2017state} regarding treatment assignment based on historical characteristics, we determine AI washing treatment status using only pre-policy data and consequently employ a classical DID framework with time-invariant assignment. Relying exclusively on pre-policy observations ensures that group classification reflects historical decoupling behavior rather than policy-induced adaptation, thereby preserving the counterfactual structure essential for causal identification. 

To measure the intensity of AI washing, we adopt the residual-based approach following previous work in earnings management through the manipulation of R\&D expenses \citep{gunny2010relation, bereskin2018real, vorst2016real}. Specifically, we estimate \autoref{eqWashing} annually during the pre-policy period using cross-sectional regressions, where residual $\varepsilon_{i,j}$ captures the component of narrative disclosure unexplained by substantive patenting output, which we interpret as AI decoupling.
\begin{align}\label{eqWashing}
    \mathrm{AI\;Word}_{i,j} = \alpha_0 + \alpha_1\mathrm{AI\;Patent}_{i,j} + \boldsymbol{\gamma}'\mathbf{Controls}_{i} + \delta_{j} + \varepsilon_{i,j}
\end{align}
where $\mathrm{AI\;Word}_{i,j}$ denotes the natural logarithm of one plus the AI-related keyword \citep{li2025impact} frequency in firm $i$'s annual report, and $\mathrm{AI\;Patent}_{i,j}$ denotes the natural logarithm of one plus the count of AI invention patents obtained independently or jointly by the firm. $\mathbf{Controls}_{i}$ absorbs the majority of firm-level heterogeneity, while $\delta_{j}$ absorbs industry fixed effects. Importantly, concerns regarding reverse causality are mitigated by the temporal structure of our variables. AI-related patent output is realized within the fiscal year, whereas textual disclosure is subsequently compiled and released in annual reports. This sequencing ensures that narrative disclosure responds to underlying innovation activities, rather than the reverse, thereby alleviating potential simultaneity concerns in estimating \autoref{eqWashing}.  

Positive residuals $\hat{\varepsilon}_{i,j}$ from \autoref{eqWashing} indicate that narrative disclosure exceeds the level attributable to substantive patenting activity, further vaildation in section \ref{External Validation} strongly support its proxy validity of AI washing. However, year-specific estimates may reflect transitory measurement noise rather than persistent decoupling behavior. To mitigate this concern, we compute the firm-level time-series average of estimated residuals ($\bar{\varepsilon}_{i}$) across the pre-policy period. We then define the treatment indicator as $\mathrm{Treat}_{i} = \mathbf{1}[\bar{\varepsilon}_{i} > 0]$.\footnote{\;$\mathbf{1}[\cdot]$ denotes the indicator function that equals one when the condition in brackets is satisfied and zero otherwise.} This classification identifies firms exhibiting systematic AI washing behavior sustained throughout the pre-treatment window, ensuring that the treatment group comprises firms with historically decoupled disclosure patterns rather than sporadic outlier observations. Appendix \ref{ApdxC} confirms that AI washing behavior exhibits substantial temporal persistence, with year-to-year rank correlation of 0.7167 at 1\% level, validating our use of multi-year averages to define treatment status.  

To validate the residual-based measure as a proxy for decoupling behavior, Appendix \ref{ApdxB} examines the pre-policy correlation between decoupling residuals and debt financing cost, establishing that firms with higher pre-policy residuals systematically enjoyed lower financing costs prior to the intervention. Furthermore, Appendix \ref{ApdxB} demonstrates that residuals exhibit significant predictive validity for next year's technological and disclosure outcomes, confirming that the measure captures persistent information risk rather than transitory measurement noise.

\subsubsection{Control Variables}

Following standard practice in corporate finance research \citep{zou2008debt, chen2009does, jin2023real, coles2023empirical, mitton2022methodological}, we include a comprehensive vector of time-varying firm-level controls. Specifically, we control for firm size (Size), financial leverage (Lev), return on assets (ROA), liquidity (Liquid), ownership concentration measured as the shareholding ratio of the top five shareholders (Top5), market valuation indicator proxied by Tobin's Q (TobinQ), and firm maturity captured by listing age (ListAge). These variables account for fundamental firm heterogeneity that may concurrently influence corporate debt financing and AI disclosure, so they are incorporated in both \autoref{eqWashing} and \autoref{eqModel}. The detail of these variables are described in Appendix \ref{ApdxA}. 

\subsection{Empirical Model}

We exploit the behavior of corporate AI washing before the release of China's 14th Five-Year Plan as a quasi-natural experiment to test our main hypotheses. Specifically, we construct the following classic DID model and expect $\beta_1$ to be positive and statistically significant.
\begin{align}\label{eqModel}
    \mathrm{Debt\;FC}_{i,t} = \beta_0 +  \beta_1\mathrm{AI\;Washing}_{i,t} + \boldsymbol{\lambda'}\mathbf{Controls}_{i,t} + \omega_i + \tau_t + \eta_{j,t} + \varphi_{k,t} + \varepsilon_{i,t}
\end{align}
where $\mathrm{AI\;Washing}_{i,t} = \mathrm{Treat}_i \times \mathrm{Post}_t$ denotes the interaction between the AI washing treatment indicator and the post-policy dummy. Based on \textbf{H1}, we expect $\beta_1 > 0$, indicating that firms with historical AI washing behavior experience a significant increase in debt financing cost following the policy implementation relative to the control group. Fixed effects are added to the regression to account for missing confounding variables. Firm ($\omega_i$) and year ($\tau_t$) fixed effects constitute the standard DID specification, capturing time-invariant firm heterogeneity and common macroeconomic shocks, respectively. We follow \citet{tan2025go} and further include industry-year ($\eta_{j,t}$) and province-year ($\varphi_{k,t}$) interaction fixed effects to address potential confounding from time-varying industry-specific trends and regional economic cycles, mitigating the risk of omitted variable bias and ensuring that the estimated treatment effect is not driven by concurrent industry or regional shocks, thereby obtaining results under a more strictly limited model to ensure robustness.

\section{Empirical Results}\label{Empirical Results}

\subsection{Descriptive Statistics}

\begin{figure}[htbp]
    \centering
    \includegraphics[width=0.75\linewidth]{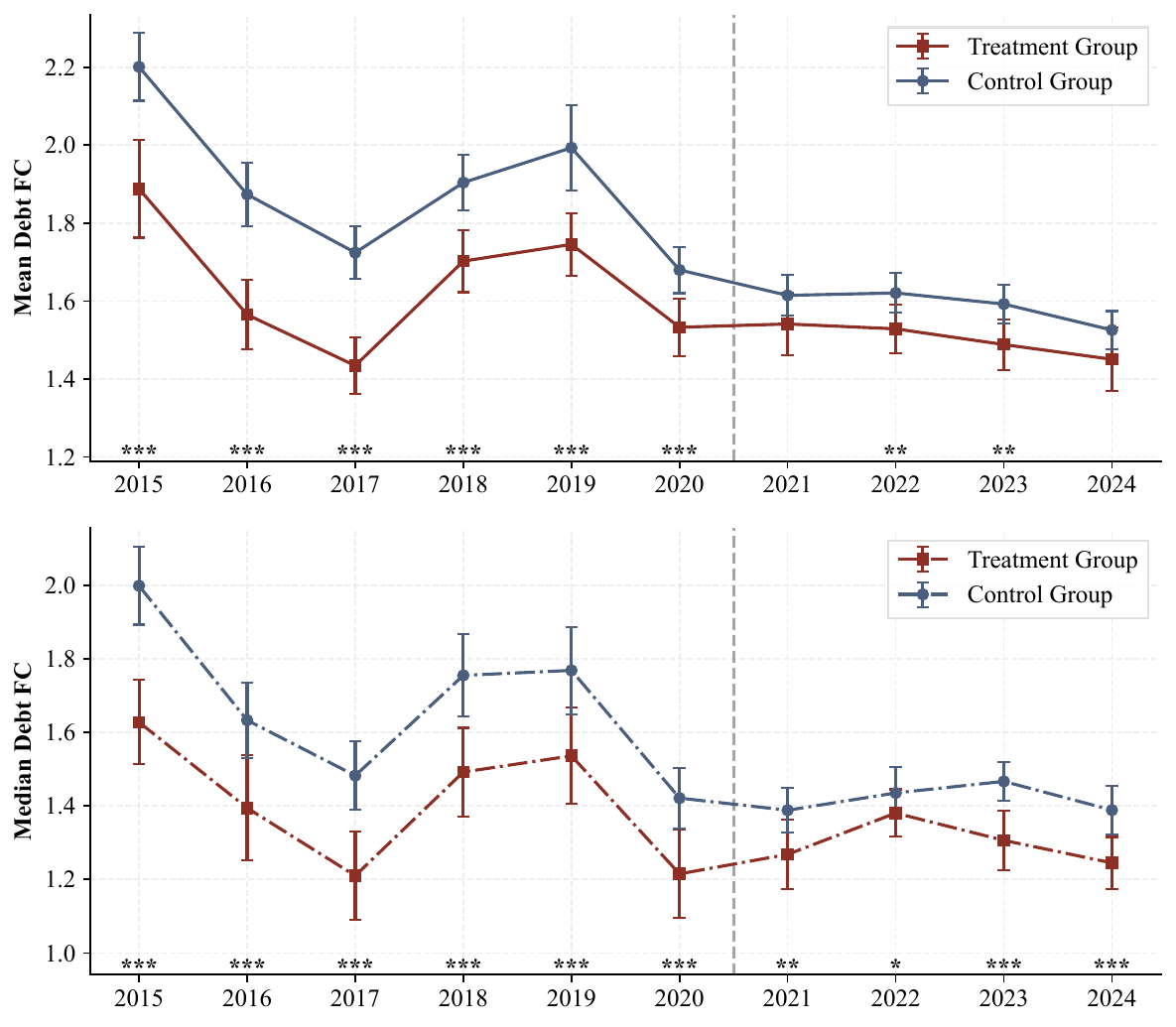}
    \caption{Line graphs of Trends in Debt Financing Cost by Treatment Status}
    \label{Debt_Financing_Cost}
\end{figure}

\autoref{tabDescriptive} presents the descriptive statistics. The sample mean of the dependent variable Debt FC is 1.6863\%, with a standard deviation of 1.5782. \autoref{Debt_Financing_Cost}\footnote{\;\autoref{Debt_Financing_Cost} Notes: (1) $^{*}$ $p<0.1$, $^{**}$ $p<0.05$, $^{***}$ $p<0.01$, for between-group difference tests. (2) Error bars represent 95\% confidence intervals.} illustrates the annual mean and median trajectory of \textit{Debt FC} for the treatment and control groups during the sample period. Visual inspection reveals that the treatment group consistently exhibits lower mean debt financing cost than the control group prior to the policy shock, with strongly significant between-group differences throughout the pre-policy window. This pattern preliminarily corroborates the financing privilege enjoyed by AI washing firms prior to the intervention. The median-based analysis mitigates the concerns regarding outlier-driven results, with treatment group medians systematically lower than those of the control group throughout 2015--2020 ($p<0.01$). Post-policy, while between-group differences persist at conventional significance levels, the magnitude of the gap narrows substantially from approximately 30--40 basis points to 10--15 basis points. Notably, the means of the two groups start to exhibit a converging trend in the year immediately preceding the policy, though between-group differences remain statistically significant, highlighting the necessity of conducting parallel trend tests to validate the pre-trends. The statistics for other control variables are comparable to those of \citet{yang2025employee} and \citet{lai2023judicial}.\footnote{\;Control variable distributions align with existing studies in terms of central tendency and dispersion, though sample size and period heterogeneity might generate minor deviations.}

\begin{table}[htbp]
	\centering
	\caption{\centering Descriptive Statistics}
    \setlength{\tabcolsep}{2pt}
    \renewcommand{\arraystretch}{1.15}
	\begin{tabular}{l*8{>{\centering\arraybackslash}p{1.7cm}}}
		\toprule
		 & $N$ & Mean & SD & Min & Med & Max & Kurtosis & Skewness \\
		\midrule
		Debt FC & 34924 & 1.6863 & 1.5782 & -20.6446 & 1.4619 & 94.6329 & 380.3777 & 7.7354 \\
        AI Washing & 34924 & 0.1690 & 0.3748 & 0.0000 & 0.0000 & 1.0000 & 4.1197 & 1.7663 \\
		Size & 34924 & 22.3536 & 1.3511 & 17.6413 & 22.1639 & 28.7908 & 4.1149 & 0.8216 \\
		Lev & 34924 & 0.4245 & 0.2094 & 0.0084 & 0.4130 & 3.5130 & 4.7194 & 0.5445 \\
		ROA & 34924 & 0.0328 & 0.0846 & -1.8591 & 0.0354 & 1.2848 & 34.5296 & -2.2109 \\
		Liquid & 34924 & 2.4959 & 2.8228 & 0.0257 & 1.6987 & 80.6637 & 87.2705 & 6.5024 \\
		Top5 & 34924 & 0.5216 & 0.1558 & 0.0081 & 0.5194 & 0.9923 & 2.5428 & 0.0501 \\
		TobinQ & 34924 & 2.1380 & 2.2164 & 0.6112 & 1.6460 & 122.1895 & 627.5802 & 17.7417 \\
		ListAge & 34924 & 2.2079 & 0.8717 & 0.0000 & 2.3026 & 3.5553 & 2.9556 & -0.7524 \\
		\bottomrule
        \addlinespace[0.1em]
        \multicolumn{9}{l}{\footnotesize Notes: N = Observations, SD = Standard Deviation, Med = Median.} \\
	\end{tabular}\label{tabDescriptive}
\end{table}

\begin{table}[htbp]
	\centering
	\caption{\centering Pearson Correlation Matrix}
    \setlength{\tabcolsep}{2pt}
    \renewcommand{\arraystretch}{1.15}
	\begin{tabular}{l*1{>{\centering\arraybackslash}p{1.47cm}}*1{>{\centering\arraybackslash}p{1.65cm}}*7{>{\centering\arraybackslash}p{1.47cm}}}
		\toprule
		 & Debt FC & AI Washing & Size & Lev & ROA & Liquid & Top5 & TobinQ & ListAge \\
		\midrule
		Debt FC & 1.000 &  &  &  &  &  &  &  &  \\
		AI Washing & -0.053*** & 1.000 &  &  &  &  &  &  &  \\
		Size & 0.125*** & 0.041*** & 1.000 &  &  &  &  &  &  \\
		Lev & 0.289*** & 0.018*** & 0.447*** & 1.000 &  &  &  &  &  \\
		ROA & -0.198*** & -0.066*** & 0.048*** & -0.378*** & 1.000 &  &  &  &  \\
		Liquid & -0.207*** & -0.007 & -0.296*** & -0.574*** & 0.183*** & 1.000 &  &  &  \\
		Top5 & -0.107*** & -0.082*** & 0.153*** & -0.065*** & 0.236*** & 0.058*** & 1.000 &  &  \\
		TobinQ & -0.078*** & -0.004 & -0.316*** & -0.134*** & 0.045*** & 0.124*** & -0.079*** & 1.000 &  \\
		ListAge & 0.152*** & 0.076*** & 0.414*** & 0.329*** & -0.222*** & -0.263*** & -0.305*** & -0.028*** & 1.000 \\
		\bottomrule
        \addlinespace[0.1em]
        \multicolumn{10}{l}{\footnotesize $^{*}$ $p<0.1$, $^{**}$ $p<0.05$, $^{***}$ $p<0.01$} \\
	\end{tabular}\label{tabCorrelation}
\end{table}

\autoref{tabCorrelation} reports the Pearson correlation coefficients. The correlations are generally moderate in magnitude, indicating that multicollinearity is unlikely to be a serious concern in this study. Moreover, consistent with the observational phenomenon of debt financing privilege, \textit{Debt FC} exhibits an overall negative correlation with AI washing.

\subsection{External Validation of AI Decoupling}\label{External Validation}

A methodological concern inherent in residual-based measures is whether the estimated decoupling reflects strategic narrative inflation or merely transitory innovation volatility. We address this concern by examining the external correlates of decoupling residuals with three observable outcomes that capture the economic and regulatory consequences of symbolic disclosure: government innovation subsidies, regulatory violations, and regulatory inquiries.

\autoref{tabExternalValidation} reports Probit estimates linking pre-policy AI decoupling residuals to subsequent external outcomes. Columns (1) and (2) reveal that firms with higher residuals are significantly more likely to receive government innovation subsidies, indicating that symbolic AI disclosure successfully extracts public resources by projecting a misleading impression of technological substance. Columns (3) and (4) show that these firms face a significantly probability of regulatory violations, confirming that the narrative facade conceals substantive misconduct rather than benign operational variance and strategic ambition. Columns (5) and (6) report contemporaneous regulatory inquiries, where the coefficient on the residual is statistically indistinguishable from zero. In sharp contrast, Column (7) reports the two-period lagged inquiry and exhibits a significant positive association. This temporal asymmetry demonstrates that AI decoupling effectively evades immediate regulatory detection, permitting the behavior to persist undetected until cumulative evidence prompts subsequent investigation. Collectively, these three dimensions establish that AI decoupling constitutes a successful deception of external stakeholders. It secures unearned resource allocations, obscures underlying violations, and eludes contemporaneous oversight, thereby validating our interpretation of the residual as a proxy for strategic AI washing.

Collectively, these findings validate the residual-based measure as capturing a strategic behavior that generates private benefits through subsidy access, invites social costs through regulatory violations, and remains partially opaque to contemporaneous oversight. This external validity supports our interpretation of the residual as a proxy for AI washing.

\begin{table}[htbp]
	\centering
	\caption{\centering External Validation of AI Decoupling}
    \setlength{\tabcolsep}{2pt}
    \renewcommand{\arraystretch}{1.15}
    \footnotesize
	\begin{tabular}{l*7{>{\centering\arraybackslash}p{1.8cm}}}
		\toprule
		 & (1) & (2) & (3) & (4) & (5) & (6) & (7) \\
		 & F.Innov\_Sub & F.Innov\_Sub & F.Violation & F.Violation & F.Inquiry & F.Inquiry & F2.Inquiry \\
		\midrule
		\multicolumn{8}{l}{\textit{\textbf{Panel A: Probit Coefficients}}} \\
        \midrule
        \addlinespace
		res & 0.114*** & 0.073*** & 0.053*** & 0.034*** & 0.020 & 0.023 & 0.062*** \\
		& (0.015) & (0.012) & (0.016) & (0.012) & (0.020) & (0.017) & (0.016) \\
		no\_patent\_tendency & $-$0.035 & 0.007 & 0.135*** & 0.101*** & 0.140*** & 0.084** & 0.162*** \\
		& (0.034) & (0.030) & (0.037) & (0.032) & (0.045) & (0.041) & (0.041) \\
		Innov\_Sub & & 1.359*** & & & & & \\
		& & (0.027) & & & & & \\
		Violation & & & & 1.557*** & & & \\
		& & & & (0.029) & & & \\
		Inquiry & & & & & & 1.566*** & 0.651*** \\
		& & & & & & (0.050) & (0.051) \\
		F.Inquiry & & & & & & & 1.408*** \\
		& & & & & & & (0.046) \\
		\midrule
		\multicolumn{8}{l}{\textit{\textbf{Panel B: Average Marginal Effects}}} \\
        \midrule
        \addlinespace
		res & 0.037*** & 0.019*** & 0.013*** & 0.007** & 0.002 & 0.002 & 0.006*** \\
		& (0.005) & (0.003) & (0.004) & (0.003) & (0.003) & (0.002) & (0.002) \\
		\midrule
		Controls & Yes & Yes & Yes & Yes & Yes & Yes & Yes \\
		Year FE & Yes & Yes & Yes & Yes & Yes & Yes & Yes \\
		Year\_Ind FE & Yes & Yes & Yes & Yes & Yes & Yes & Yes \\
		Year\_Prov FE & Yes & Yes & Yes & Yes & Yes & Yes & Yes \\
		$N$ & 19179 & 19179 & 19121 & 19121 & 18805 & 18805 & 18377 \\
		Pseudo $R^{2}$ & 0.153 & 0.299 & 0.075 & 0.262 & 0.200 & 0.320 & 0.316 \\
		\bottomrule
        \addlinespace[0.1em]
        \multicolumn{8}{l}{\footnotesize * $p<0.1$, ** $p<0.05$, *** $p<0.01$} \\
        \multicolumn{8}{l}{\footnotesize Note: Panel A reports Probit coefficients. Panel B reports average marginal effects ($dy/dx$).} \\
        \multicolumn{8}{l}{\footnotesize Standard errors clustered at firm level in parentheses.} \\
        \multicolumn{8}{l}{\footnotesize $no\_patent\_tendency$ is a dummy variable equal to one if the firm has accumulated zero AI patents} \\
        \multicolumn{8}{l}{\footnotesize in the current observation year, indicating control for firms that have never produced any patents.} \\
	\end{tabular}\label{tabExternalValidation}
\end{table}

\subsection{Baseline Results}

\autoref{tabBaseline} reports the differences in debt financing between firms classified as AI-washing and non-AI-washing by our methodology. Columns (1) to (4) present the results with progressively more stringent fixed effects specifications. Column (1) includes only firm and year fixed effects. Column (2) augment the specification with firm-level control variables. Columns (3) to (4) further add industry-year and province-year interaction fixed effects, respectively, to account for time-varying industry-specific and region-specific shocks.

\begin{table}[htbp]
	\centering
	\caption{\centering Baseline Results}
    \setlength{\tabcolsep}{2pt}
    \renewcommand{\arraystretch}{1.15}
    \footnotesize
	\begin{tabular}{l*4{>{\centering\arraybackslash}p{3.45cm}}}
		\toprule
		 & (1) & (2) & (3) & (4) \\
		 & Debt FC & Debt FC & Debt FC & Debt FC \\
		\midrule
		AI Washing & 0.193*** & 0.144*** & 0.136*** & 0.125*** \\
		& (0.037) & (0.035) & (0.036) & (0.036) \\
		Size &  & -0.045 & -0.034 & -0.030 \\
		&  & (0.056) & (0.056) & (0.056) \\
		Lev &  & 1.164*** & 1.190*** & 1.182*** \\
		&  & (0.146) & (0.143) & (0.145) \\
		ROA &  & -1.258*** & -1.140*** & -1.112*** \\
		&  & (0.228) & (0.229) & (0.226) \\
		Liquid &  & 0.007 & 0.008 & 0.008 \\
		&  & (0.011) & (0.011) & (0.011) \\
		Top5 &  & -0.851*** & -0.871*** & -0.839*** \\
		&  & (0.170) & (0.167) & (0.167) \\
		TobinQ &  & -0.008 & -0.008 & -0.007 \\
		&  & (0.016) & (0.015) & (0.015) \\
		ListAge &  & 0.192*** & 0.210*** & 0.196*** \\
		&  & (0.038) & (0.040) & (0.041) \\
		\_cons & 1.654*** & 2.232* & 1.934* & 1.875 \\
		& (0.006) & (1.186) & (1.170) & (1.174) \\
        \midrule
		Firm FE & Yes & Yes & Yes & Yes \\
		Year FE & Yes & Yes & Yes & Yes \\
		Year\_Ind FE &  &  & Yes & Yes \\
		Year\_Prov\_FE &  &  &  & Yes \\
		$N$ & 34924 & 34924 & 34924 & 34924 \\
		Adj.\;$R^{2}$ & 0.474 & 0.489 & 0.495 & 0.498 \\
		\bottomrule
        \addlinespace[0.1em]
        \multicolumn{5}{l}{\footnotesize $^{*}$ $p<0.1$, $^{**}$ $p<0.05$, $^{***}$ $p<0.01$} \\
        \multicolumn{5}{l}{\footnotesize Note: Standard errors clustered at firm level in parentheses.} \\
	\end{tabular}\label{tabBaseline}
\end{table}

Column (2) employs a two-way fixed-effects model with control variables. The coefficient on \textit{AI Washing} is 0.144 and significant positive at the 1\% level, indicating that firms with historical AI washing behavior experience a 14.4 basis points greater increase in debt financing cost relative to genuine AI adopters following the policy shock. This magnitude represents approximately 8.5\% of the sample mean Debt FC (1.69), constituting an economically meaningful differential cost increase. Control variable coefficients align with standard corporate finance predictions. Financial leverage (\textit{Lev}) elevate debt financing cost \citep{baxter1967leverage} whereas profitability (\textit{ROA}) and ownership concentration (\textit{Top5}) reduce them \citep{jiang2008beating, sanchez2011ownership}. The positive coefficient on \textit{ListAge} likely reflects that mature firms have exhausted low-cost debt capacity and must accept higher marginal financing costs for subsequent borrowing \citep{vanacker2010pecking}.

As we progressively introduce more stringent fixed effects controlling for concurrent industry and regional trends, the coefficient on AI washing attenuates modestly from 0.144 in Column (2) to 0.136 in Column (3) and 0.125 in Column (4), yet remains statistically significant at the 1\% level. This stability confirms that the effect is robust to potential omitted variable bias from unobserved heterogeneity at the industry or regional level. Our preferred baseline specification in Column (4) indicates that AI washing firms experience a 12.5 basis points greater increase in debt financing cost relative to the control group. This magnitude is comparable to the cost increases associated with regulatory penalties documented in prior literature \citep{gong2021punishment}, thereby strongly supporting \textbf{H1}. The adjusted $R^2$ of 0.498 indicates that the full specification explains approximately 50\% of the variation in debt financing cost while retaining adequate identifying variation despite the inclusion of high-dimensional fixed effects.

\subsection{Robustness Tests}

\subsubsection{Addressing Observable Sample Selection Bias}

AI washing firms may differ systematically from genuine adopters in observable characteristics, potentially violating the parallel trend assumption. To address this observable sample selection bias, we follow \citet{xie2025does} and implement two reweighting procedures.

First, we employ Propensity Score Matching (PSM) to construct a counterfactual sample \citep{rosenbaum1983central} wherein each AI washing firms is matched with two non-washing counterparts exhibiting similar observable characteristics. We estimate propensity scores via logit regression including all baseline controls. To balance matching quality against sample retention, we perform 1:2 nearest-neighbor matching with replacement within a caliper of 0.01. Second, we implement Entropy Balancing (EB) as an alternative reweighting scheme that achieves exact first-moment balance on the covariate distribution \citep{hainmueller2012entropy}. We conduct EB annually for the pre-treatment period (2015--2020) and apply the mean weights to the full sample (2015--2024), thereby exploiting pre-treatment covariate dynamics while maintaining procedure comparability with PSM.

\begin{figure}[htbp]
    \centering
    \includegraphics[width=\textwidth]{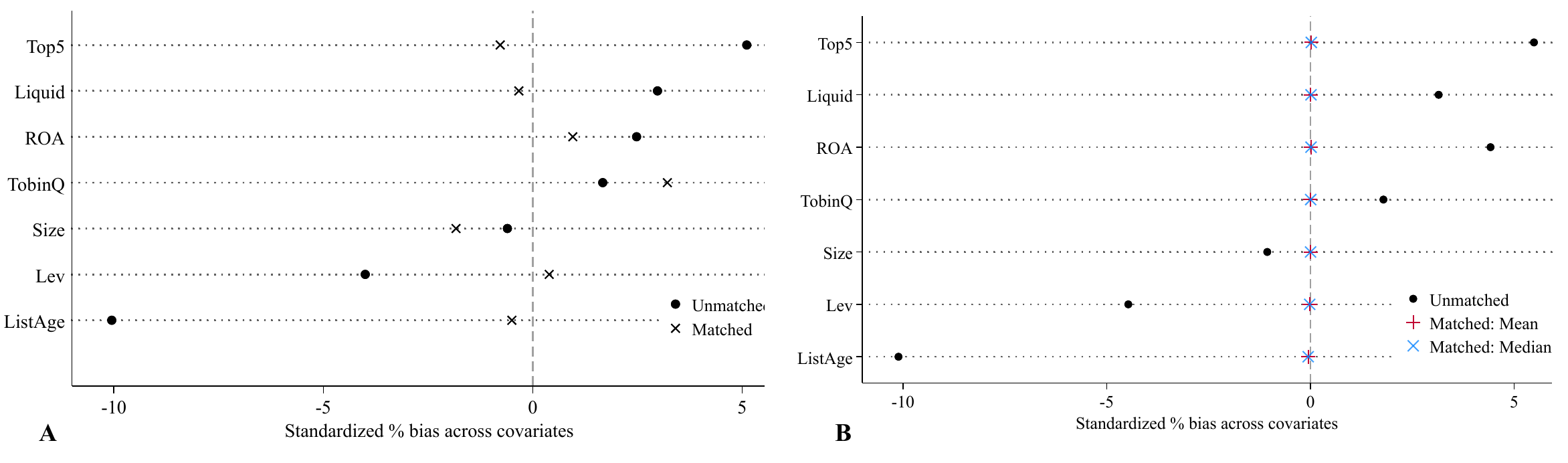}
    \caption{Covariate Balance Diagnostics of PSM and EB}
    \label{figMatch}
\end{figure}

\autoref{figMatch} presents covariate balance diagnostics. Panel A shows that PSM reduces standardized biases to within $\pm5\%$, but imbalances remain for \textit{TobinQ} and \textit{Size}. This motivates the complementary use of Entropy Balancing. Panel B confirms that EB achieves exact mean balance, with all standardized biases eliminated. These diagnostics confirm that both methodologies eliminate observable selection bias, with EB providing more precise distributional alignment without sample attrition.

\autoref{tabRobustnessPSMEB} reports the re-estimation results. Columns (1) and (2) report estimates using the PSM-matched sample. The coefficient on AI washing remains significant at the 1\% level at 0.122, comparable in magnitude to the baseline estimate of 0.125. Columns (3) and (4) report EB-weighted estimates, with the coefficient remaining positive and significant at 0.117. Consistency across matching methodologies alleviates concerns that the documented narrowing gap of \textit{Debt FC} merely reflects pre-treatment observable differences between groups.

\begin{table}[htbp]
	\centering
	\caption{\centering Propensity Score Matching and Entropy Balance Results}
    \setlength{\tabcolsep}{2pt}
    \renewcommand{\arraystretch}{1.15}
    \footnotesize
	\begin{tabular}{l*4{>{\centering\arraybackslash}p{3.45cm}}}
		\toprule
		 & \multicolumn{2}{c}{PSM} & \multicolumn{2}{c}{EB} \\
        \cmidrule(r){2-3} \cmidrule(r){4-5}
		 & Debt FC & Debt FC & Debt FC & Debt FC \\
		\midrule
		AI Washing & 0.139*** & 0.122*** & 0.134*** & 0.117*** \\
		& (0.035) & (0.036) & (0.035) & (0.035) \\
		Size & -0.044 & -0.029 & -0.068 & -0.059 \\
		& (0.057) & (0.057) & (0.061) & (0.060) \\
		Lev & 1.161*** & 1.180*** & 1.138*** & 1.163*** \\
		& (0.147) & (0.146) & (0.145) & (0.144) \\
		ROA & -1.269*** & -1.126*** & -1.276*** & -1.144*** \\
		& (0.232) & (0.229) & (0.202) & (0.204) \\
		Liquid & 0.007 & 0.008 & 0.010 & 0.011 \\
		& (0.011) & (0.012) & (0.012) & (0.013) \\
		Top5 & -0.839*** & -0.833*** & -0.929*** & -0.892*** \\
		& (0.171) & (0.168) & (0.176) & (0.172) \\
		TobinQ & -0.007 & -0.006 & -0.014 & -0.012 \\
		& (0.018) & (0.017) & (0.012) & (0.012) \\
		ListAge & 0.198*** & 0.201*** & 0.201*** & 0.205*** \\
		& (0.039) & (0.042) & (0.036) & (0.039) \\
		\_cons & 2.190* & 1.826 & 2.756** & 2.508* \\
		& (1.206) & (1.193) & (1.308) & (1.281) \\
        \midrule
		Firm FE & Yes & Yes & Yes & Yes \\
		Year FE & Yes & Yes & Yes & Yes \\
		Year\_Ind FE &  & Yes &  & Yes \\
		Year\_Prov FE &  & Yes &  & Yes \\
		$N$ & 34386 & 34386 & 34924 & 34924 \\
		Adj.\;$R^{2}$ & 0.488 & 0.497 & 0.491 & 0.500 \\
		\bottomrule
        \addlinespace[0.1em]
        \multicolumn{5}{l}{\footnotesize $^{*}$ $p<0.1$, $^{**}$ $p<0.05$, $^{***}$ $p<0.01$} \\
        \multicolumn{5}{l}{\footnotesize Notes: (1) Mean EB-weights of sample 2015-2020 applied to full sample 2015-2024.} \\
        \multicolumn{5}{l}{\footnotesize (2) Standard errors clustered at firm level in parentheses.}
	\end{tabular}\label{tabRobustnessPSMEB}
\end{table}

\subsubsection{Parallel Trend Test}

\begin{figure}[htbp]
    \centering
    \includegraphics[width=0.7\textwidth]{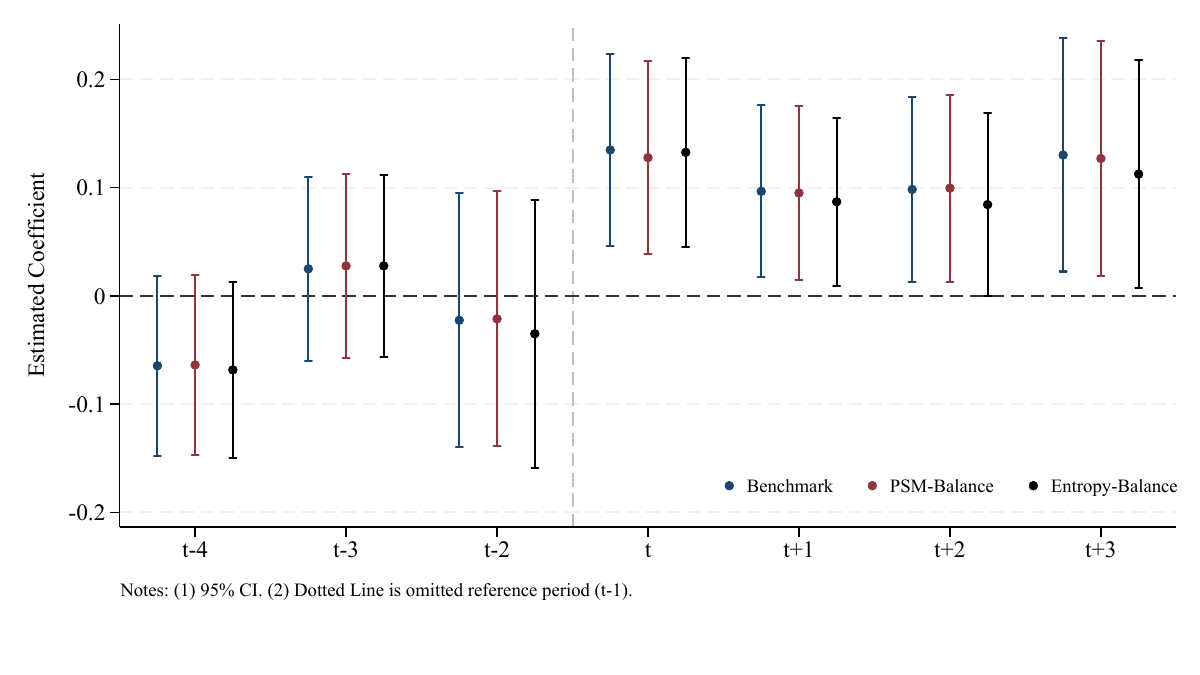}
    \caption{Parallel Trend Test}
    \label{figParallelTrend}
\end{figure}

The DID identification strategy requires parallel pretreatment trends in debt financing cost between treatment and control groups. We test this assumption using the event-study specification in \autoref{eqEvent} where $\mathcal{T} = \{-4,-3,-2,0,1,2,3\}$ denotes event-time indicators relative to the 2021 policy announcement and $\tau=-1$ serves as the omitted reference period. The coefficients $\beta_{\tau}$ capture differential debt financing cost between treatment and control groups in period $\tau$ with the null hypothesis of parallel trends requiring $\beta_{\tau}=0$ for all $\tau<0$.

\begin{align}\label{eqEvent}
\mathrm{Debt\;FC}_{i,t} = \sum_{\tau \in \mathcal{T}} \beta_{\tau} (\mathrm{Treat}_i \times \mathbf{1}[t=2021+\tau]) + \boldsymbol{\lambda}'\mathbf{Controls}_{i,t} + \omega_i + \tau_t +  \eta_{j,t} + \varphi_{k,t} + \varepsilon_{i,t}
\end{align}

\autoref{figParallelTrend} plots the estimated coefficients and 95\% confidence intervals across the baseline, PSM-matched, and entropy-balanced samples. For the pre-treatment periods ($\tau = -4$ to $-2$), all point estimates cluster around zero with confidence intervals spanning zero, indicating no pretreatment divergence in \textit{Debt FC} between groups. This confirms the parallel trends assumption. In the post-treatment periods ($\tau = 0$ to $3$), \textit{Debt FC} increases significantly, suggesting persistent policy effects on debt market pricing. At $\tau = 2$, the entropy-balanced estimate shows confidence intervals marginally including zero, consistent with modest attenuation of the treatment effect over time.

\subsubsection{Placebo Permutation Test}

To address concerns that \autoref{eqModel} estimates might reflect random noise rather than actual policy impact, we conduct a placebo permutation test following \citet{abadie2010synthetic}. We generate 1000 pseudo-samples by randomly shuffling the treatment indicator among firms. For each permutation we estimate the baseline regression specification and store the coefficient on the pseudo-treatment variable.

\begin{figure}[htbp]
    \centering
    \includegraphics[width=0.7\textwidth]{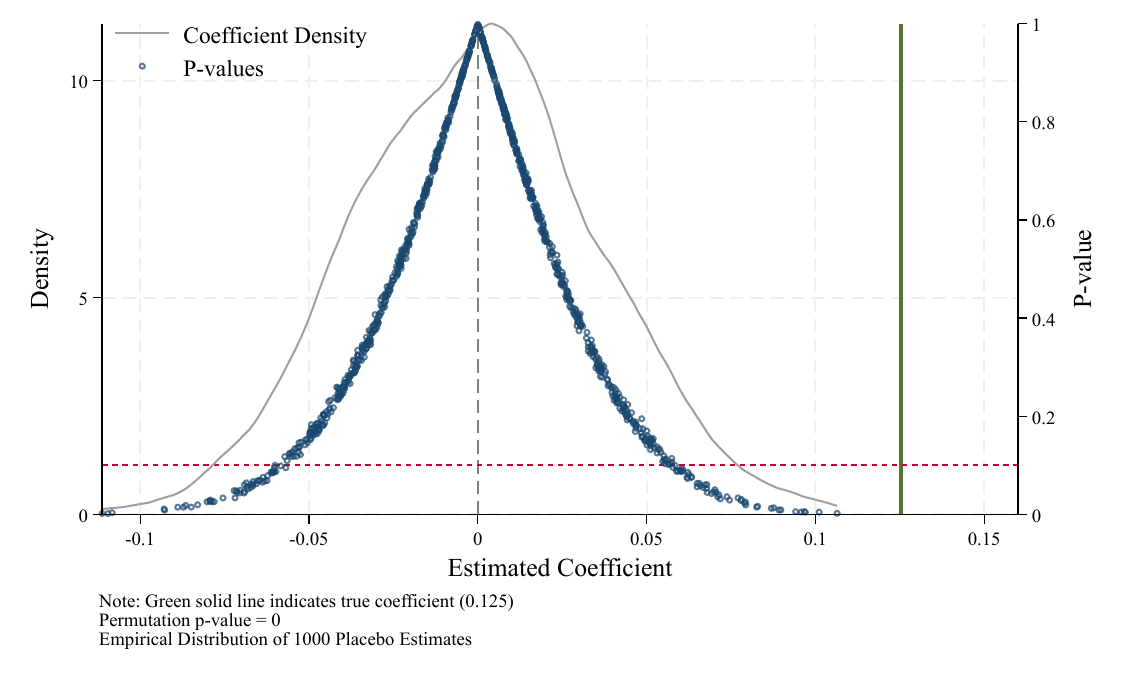}
    \caption{Placebo Permutation Test}
    \label{figPlacebo}
\end{figure}

\autoref{figPlacebo} plots the kernel density of placebo coefficients and their associated $p$-values. The placebo coefficients concentrate tightly around zero and their $p$-values cluster above conventional significance thresholds, indicating that randomly assigned treatments yield statistically insignificant effects consistent with the null hypothesis. The true coefficient of 0.125 lies substantially outside the placebo distribution. None of the 1000 randomizations produces a coefficient as extreme as the actual estimate, confirming that the observed relative increase in debt financing cost for AI washing firms is attributable to the policy shock rather than random data fluctuations or unobserved confounding structures, thereby validating the causal interpretation of the baseline findings.

\subsubsection{Industry Subsamples and Alternative Treatment Indicators}

To verify that the baseline findings are not sensitive to sample composition or the specific measurement approach used to identify AI washing, we conduct two sets of robustness checks. First, we exclude firms in CSRC Sector I (Information transmission, software and information technology services), which encompasses telecommunications (I63), internet and related services (I64), and software and IT services (I65). Firms in these industries frequently protect AI capabilities via trade secrets, software copyrights, or non-patentable algorithms, rendering patent counts potentially uninformative about true AI capability. As reported in \autoref{tabChangeMeasurement}, Column (1), the coefficient after dropping this sector is 0.110 and remains significant at the 1\% level, closely tracking the magnitudes in the remaining columns. This alleviates concerns that the documented financing cost penalty is driven by cross-industry differences in patenting propensity.

Second, we construct four alternative treatment indicators using the full sample. First, we employ the stock of AI patents rather than the annual flow when estimating \autoref{eqWashing}. By aggregating patenting outcomes over time, this measure avoids misclassifying firms as AI washers due to temporary declines in annual patent output that reflect innovation volatility rather than intentional decoupling. Second, we adopt the standardized difference approach proposed by \citet{yu2020greenwashing} and construct an alternative measure that captures the divergence between annual AI narrative disclosures and contemporaneous AI patent stocks after adjusting for industry-specific means and variances:
\begin{align}\label{eqGreenwashing}
    \mathrm{Z}_{i,j}^{\mathrm{AI}}=\dfrac{\left(\mathrm{AI\;Word}_{i,j}-\overline{\mathrm{AI\;Word}_{j}}\right)}{\sigma^{\mathrm{AI\;Word}}_{i,j}} - \dfrac{\left(\mathrm{AI\;Patent}_{i,j}-\overline{\mathrm{AI\;Patent}_{j}}\right)}{\sigma^{\mathrm{AI\;Patent}}_{i,j}}
\end{align}
Third, we restrict the residual calculation to the fiscal year immediately preceding the policy announcement (2020), identifying treatment status based solely on decoupling behavior observed at the policy threshold. Fourth, we replace granted patents with AI patent applications to construct the decoupling residual. Because patent applications reflect contemporaneous innovation investment decisions with minimal administrative lag, this specification mitigates concerns that our baseline captures merely pre-policy innovation stocks rather than post-shock capability accumulation.

\autoref{tabChangeMeasurement} reports the results. Column (2) employs AI patent stock rather than flow, yielding a coefficient of 0.110 that remains significant at the 1\% level. This confirms that the documented dissipation of financing advantages for AI washing firms persists when cumulative technological capabilities anchor the decoupling calculation, ruling out that transitory fluctuations in annual patenting drive the results. Column (3) reports estimates using the standardized difference measure, with a coefficient of 0.113 that closely tracks the Column (1) magnitude, validating that the residual-based approach and the standardized divergence metric capture similar underlying decoupling behavior. Column (4) restricts the decoupling calculation to the 2020 cross-section, producing a coefficient of 0.081 that is significant at the 5\% level. The modest attenuation in this specification reflects increased measurement noise from single-year classification, yet the persistent significance confirms that firms exhibiting decoupling behavior immediately prior to the policy shock indeed experience a relative increase in debt financing cost. Column (5) replaces granted patents with AI patent applications; the coefficient of 0.101 remains statistically significant at the 1\% level and economically comparable to the other specifications, confirming that the documented financing cost penalty is not driven by patent examination delays. This specification also addresses the alternative interpretation that our treatment group comprises early-stage innovators or firms acquiring AI capabilities externally: if such firms were genuinely ramping up AI investment during the post-policy window, rational intellectual property strategies would likely generate observable patent applications. Moreover, because firms relying on trade-secret protection typically reduce public technical disclosure to avoid leakage, the coexistence of dense AI narratives and low patent applications is inconsistent with the trade-secret hypothesis.

The consistency of results across these alternative sample constructions and measurement approaches alleviates concerns that the identified dissipation of financing advantages is sensitive to the particular measurement approach or sample composition used to identify AI washing.

\begin{table}[htbp]
	\centering
	\caption{\centering Robustness to Industry Subsamples and Alternative Treatment Definitions}
    \setlength{\tabcolsep}{2pt}
    \renewcommand{\arraystretch}{1.15}
    \footnotesize
	\begin{tabular}{l*5{>{\centering\arraybackslash}p{2.72cm}}}
		\toprule
      & Exclude IT Industry & \multicolumn{3}{c}{Alternative Measurements} \\
      \cmidrule(r){2-2}  \cmidrule(r){3-6}
		 & Debt FC & Debt FC & Debt FC & Debt FC & Debt FC \\
		\midrule
		AI Washing & 0.110*** &  &  &  &\\
		& (0.038) &  &  & & \\
		AI Washing\_1 &  & 0.110*** &  & & \\
		&  & (0.035) &  &  &\\
		AI Washing\_2 &  &  & 0.113*** & & \\
		&  &  & (0.037) & & \\
		AI Washing\_3 &  &  &  & 0.081**& \\
		&  &  &  & (0.034) &\\
        AI Washing\_4 & & & & & 0.101*** \\
        & & & & & (0.035) \\
        \midrule
        Controls & Yes & Yes & Yes & Yes & Yes \\
		Firm FE & Yes & Yes & Yes & Yes & Yes \\
		Year FE & Yes & Yes & Yes & Yes & Yes \\
		Year\_Ind FE & Yes & Yes & Yes & Yes & Yes \\
		Year\_Prov FE & Yes & Yes & Yes & Yes & Yes \\
		$N$ & 32164 & 34924 & 34924 & 33460 & 34924 \\
		Adj.\;$R^{2}$ & 0.489 & 0.498 & 0.498 & 0.501 & 0.498 \\
		\bottomrule
        \addlinespace[0.1em]
        \multicolumn{6}{l}{\footnotesize $^{*}$ $p<0.1$, $^{**}$ $p<0.05$, $^{***}$ $p<0.01$} \\
        \multicolumn{6}{l}{\footnotesize Note: Standard errors clustered at firm level in parentheses.} \\
	\end{tabular}\label{tabChangeMeasurement}
\end{table}

The binary treatment classification may obscure heterogeneous effects across the severity of AI washing. To examine whether the relative increase in debt financing cost varies with the magnitude of pre-policy AI washing, we implement intensity DID specifications that exploit continuous variation in the residuals.

\begin{table}[htbp]
	\centering
	\caption{\centering Treatment Intensity and Dose-Response Effects}
    \setlength{\tabcolsep}{2pt}
    \renewcommand{\arraystretch}{1.15}
    \footnotesize
	\begin{tabular}{l*4{>{\centering\arraybackslash}p{3.4cm}}}
		\toprule
      & \multicolumn{2}{c}{Intensity DID} & \multicolumn{2}{c}{Quantile DID} \\
     \cmidrule(r){2-3} \cmidrule(r){4-5}
		 & Debt FC & Debt FC & Debt FC & Debt FC \\
		\midrule
		AI Washing\_I1 & 0.070*** &  &  &  \\
		& (0.020) &  &  &  \\
		AI Washing\_I2 &  & 0.058*** &  &  \\
		&  & (0.017) &  &  \\
		AI Washing\_Q2 &  &  & 0.100** & 0.073 \\
		&  &  & (0.045) & (0.054) \\
		AI Washing\_Q3 &  &  & 0.151*** & 0.125*** \\
		&  &  & (0.044) & (0.047) \\
		AI Washing\_Q4 &  &  &  & 0.180*** \\
		&  &  &  & (0.053) \\
        \midrule
        Controls & Yes & Yes & Yes & Yes \\
		Firm FE & Yes & Yes & Yes & Yes \\
		Year FE & Yes & Yes & Yes & Yes \\
		Year\_Ind FE & Yes & Yes & Yes & Yes \\
		Year\_Prov FE & Yes & Yes & Yes & Yes \\
		$N$ & 34924 & 34924 & 34924 & 34924 \\
		Adj.\;$R^{2}$ & 0.498 & 0.498 & 0.498 & 0.498 \\
		\bottomrule
        \addlinespace[0.1em]
        \multicolumn{5}{l}{\footnotesize * $p<0.1$, ** $p<0.05$, *** $p<0.01$} \\
        \multicolumn{5}{l}{\footnotesize Notes: (A) Columns (1)-(2): intensity DID with raw and standardized $\hat{\varepsilon}_{i,j}$.} \\
        \multicolumn{5}{l}{\footnotesize (B) Columns (3): Q2 \& Q3 are median split of $\hat{\varepsilon}_{i,j}>0$ group (Q1: $\hat{\varepsilon}_{i,j}\leq0$).} \\
        \multicolumn{5}{l}{\footnotesize (C) Columns (4): Q2 to Q4 are terciles of $\hat{\varepsilon}_{i,j}>0$ group (Q1: $\hat{\varepsilon}_{i,j}\leq0$).} \\
        \multicolumn{5}{l}{\footnotesize (D) Standard errors clustered at firm level in parentheses.} \\
	\end{tabular}\label{tabIntensity}
\end{table}

We exploit continuous variation in decoupling intensity to estimate dose-response relationships. Columns (1) and (2) report intensity DID estimates using the raw residual and its standardized transformation. The coefficient of 0.070 indicates that each one-unit increase in the pre-policy decoupling residual elevates relative debt financing cost by 0.07 percentage points post-policy, representing a 4.1\% proportional increase relative to the sample mean of 1.69\%. The standardized coefficient of 0.058 implies that a one-standard-deviation increase in decoupling intensity generates 0.058 percentage points higher debt financing cost, equivalent to 3.7\% of the standard deviation or 3.4\% of the mean level. Both estimates are significant at the 1\% level, confirming that firms with more severe pre-policy AI washing experience proportionally larger increases in debt financing cost and that the documented dissipation of financing advantage intensifies with decoupling severity.

Columns (3) and (4) examine nonlinearities by partitioning AI washing firms into quantiles based on decoupling intensity while retaining non-washing firms as the base group. Column (3) reports a median split of the positive residual group (median-split of $\hat{\varepsilon}_{i,j}>0$), yielding coefficients of 0.100 and 0.151 for the lower and upper halves respectively. The monotonic pattern indicates that even within the population of AI washing firms, those with more severe decoupling face disproportionately higher debt financing cost. Column (4) refines this analysis using a tercile split of the positive residuals, producing coefficients of 0.073, 0.125, and 0.180. Notably, the lowest tercile is statistically indistinguishable from zero whereas the middle and upper terciles exhibit significant and strictly increasing penalties. This pattern reveals that mild AI washing does not trigger significant creditor behavior responses whereas moderate to severe AI washing generates economically meaningful increases in debt financing cost. This approach validates that the baseline results are not artifacts of specific treatment construction choices.

\subsubsection{Addressing Selection from Unobservable Factors and Post-Policy Entry}

To correct for sample selection bias arising from the nonrandom restriction to firms with observable AI activities, we employ the Heckman two-stage selection model. Our baseline sample conditions on firms exhibiting positive AI disclosure or patenting during 2015--2020, potentially inducing bias if unobservable characteristics simultaneously determine sample inclusion and debt financing cost.

The first stage estimates the probability of exhibiting observable AI activities using Probit on the full sample of listed firms including post-2021 entrants with 39016 observations. We employ \textit{IT\_ratio} as an instrument. Higher technology executive ratios reduce barriers to formulating AI narratives \citep{jiang2024does} and increase disclosure likelihood when patent stocks are insufficient, yet do not alter firm fundamentals such as asset quality or cash flow volatility \citep{gore2011role, czarnitzki2009capital}, satisfying the exclusion restriction. We control for \textit{no\_entry} to account for structural truncation, as firms listed after 2021 lack the requisite longitudinal history for computing pre-treatment decoupling residuals by construction.

The second stage estimates the debt financing cost equation on the incumbent subsample with 34924 observations using OLS and incorporates the inverse Mills ratio to correct for potential selection bias. Under the null hypothesis of no selection bias, the \textit{IMR} coefficient should be statistically indistinguishable from zero, indicating that unobservable factors determining sample inclusion do not systematically correlate with debt financing cost conditional on observables.

\autoref{tabHeckman} reports the results. In the first stage, \textit{IT\_ratio} is 0.982 and significant at the 1\% level, confirming that technology-intensive management teams are more likely to enter the sample. The \textit{no\_entry} coefficient is -0.164 and significant at the 1\% level. This validates the structural truncation of the sample, as post-2021 entrants are absent from the pre-policy window by construction rather than by random selection, ensuring that the selection mechanism is governed by entry timing relative to the policy threshold. Controlling for this composition discontinuity purges the identification of the IT intensity effect from survivorship bias arising from entry timing. In the second stage, the coefficient on \textit{AI Washing} remains 0.124 ($p<0.01$), while the \textit{IMR} is insignificant ($p=0.356$), indicating that unobservable factors determining sample inclusion are orthogonal to debt pricing conditional on observables. The robustness of the AI washing effect across the Heckman correction validates that our findings are not driven by selection on unobserved firm heterogeneity.

\begin{table}[htbp]
	\centering
	\caption{\centering Heckman Two-Stage Selection Model}
    \setlength{\tabcolsep}{2pt}
    \renewcommand{\arraystretch}{1.15}
    \footnotesize
	\begin{tabular}{l*2{>{\centering\arraybackslash}p{3.5cm}}}
		\toprule
		 & (1) & (2) \\
		 & First Stage (Probit) & Second Stage (OLS) \\
		\midrule
		AI Washing &  & 0.124*** \\
		&  & (0.036) \\
		IMR &  & 0.155 \\
		&  & (0.168) \\
		IT\_ratio & 0.982*** &  \\
		& (0.079) &  \\
        \midrule
		no\_entry & -0.164*** &  \\
		& (0.049) &  \\
		Size & 0.207*** & -0.015 \\
		& (0.014) & (0.054) \\
		Lev & -0.226*** & 1.167*** \\
		& (0.083) & (0.143) \\
		ROA & -0.150 & -1.123*** \\
		& (0.135) & (0.223) \\
		Liquid & -0.002 & 0.008 \\
		& (0.004) & (0.011) \\
		Top5 & -0.122 & -0.845*** \\
		& (0.096) & (0.167) \\
		TobinQ & 0.022*** & -0.005 \\
		& (0.006) & (0.015) \\
		ListAge & -0.164*** & 0.181*** \\
		& (0.019) & (0.040) \\
        \midrule
		\multirow{2}{*}{FE} & \multirow{2}{*}{Year, Ind} & Firm, Year, Year\_Ind, Year\_Prov \\
		P-value: \textit{IMR}=0 &  & 0.356 \\
		$N$ & 39016 & 34924 \\
		Pseudo/Adj.\;$R^{2}$ & 0.169 & 0.498 \\
		\bottomrule
        \addlinespace[0.1em]
        \multicolumn{3}{l}{\footnotesize $^{*}$ $p<0.1$, $^{**}$ $p<0.05$, $^{***}$ $p<0.01$} \\
        \multicolumn{3}{l}{\footnotesize Notes: (1) \textit{IMR} is insignificant, indicating no severe sample selection bias.} \\
        \multicolumn{3}{l}{\footnotesize (2) Standard errors clustered at firm level in parentheses.} \\
	\end{tabular}\label{tabHeckman}
\end{table}

\subsection{Excluding Competitive Policy Effects}

Previous robustness tests establish that benchmark estimates are insensitive to observable and unobservable sample selection biases and to alternative treatment group constructions. Nevertheless, the 2021 policy window coincides with an exceptionally high density of regulatory interventions that may independently influence corporate debt financing cost. To ensure that the documented relative increase in debt financing cost for AI washing firms is attributable specifically to China's 14th Five-Year Plan rather than confounding policy contemporaneities, we systematically examine the policy landscape during 2021 through 2024.

We explicitly control for three major interventions representing the most salient regulatory shocks during the sample window. The ``SRDI Little Giant'' certification program operates through firm-specific recognition status that directly alters bank lending terms, presenting a confounding factor that varies within industry-year cells. The real estate ``Three Red Lines'' policy and the green credit policy generate cross-sectional variation across industries and years, yet we explicitly account for them to address potential sector-specific debt repricing and bank credit reallocation effects that may spill over into non-targeted sectors through general equilibrium adjustments.

\begin{table}[htbp]
	\centering
	\caption{\centering Competitive Policy Interventions and Exclusion Justifications}
    \setlength{\tabcolsep}{4pt}
    \renewcommand{\arraystretch}{1.15}
	\begin{tabular}{l*1{>{\arraybackslash}p{5.2cm}} *1{>{\arraybackslash}p{5.8cm}}}
		\toprule
		\textbf{Policy Intervention} & \textbf{Description} & \textbf{Impact Channel / Exclusion Rationale} \\
		\midrule
		\multicolumn{3}{l}{\textit{\textbf{Included in Robustness Checks}}} \\
		\addlinespace
		``SRDI Little Giant'' Certification & Official certification conferring preferential credit access and regulatory forbearance & Reduces debt financing cost and expands credit availability through certified status \\
		\addlinespace
		``Three Red Lines'' Policy & Imposes stringent balance sheet constraints on real estate developers & Tightens sector-specific credit supply and generates pricing spillovers via resource reallocation to other sectors \\
		\addlinespace
		Green Credit Policy & Differential financing treatment contingent based on environmental performance & Alters debt cost structure through subsidized rates for compliant firms and credit rationing for polluting firms \\
		\midrule
		\multicolumn{3}{l}{\textit{\textbf{Excluded from Explicit Checks}}} \\
		\addlinespace
		BSE Establishment & Dedicated equity financing venue for innovative small and medium firms & Primarily affects equity valuation, and our sample requires the firms have been listed pre-policy. \\
		\addlinespace
		Platform Economy Antitrust & Operational compliance mandates targeting internet platform concentration & Limited to internet giants and related to operations, not directly related to debt financing cost \\
		\addlinespace
		R\&D Expense Super-deduction & Tax expenditure incentives for qualified research and development investments & Affects corporate tax burden and cash flow timing rather than debt financing cost \\
		\bottomrule
	\end{tabular}\label{tabCompetitivePolicies}
\end{table}

Other contemporaneous initiatives listed in the lower panel of \autoref{tabCompetitivePolicies} are either absorbed by the high-dimensional fixed effects structure or excluded from explicit robustness checks because their primary impact channels do not directly affect debt contracting terms. The Beijing Stock Exchange (BSE) establishment targets equity financing channels rather than bank debt markets. The platform economy antitrust regulations affect a limited subset of internet giants in our sample and operate primarily through equity valuation and operational compliance costs. The R\&D expense super-deduction influences corporate tax burdens rather than debt pricing directly. These exclusion criteria ensure that our robustness checks focus on the most relevant confounding factors while maintaining empirical tractability.

\subsubsection{Excluding ``SRDI Little Giant'' Certification}

First, we address potential confounding from the ``SRDI Little Giant'' certification. This designation directly reduces debt financing cost for certified firms through preferential lending terms and regulatory forbearance. Because the program targets technology-intensive firms and overlaps temporally with our treatment window, it may independently influence the debt cost structure of AI washing firms.\footnote{\;``SRDI'' denotes the official classification of firms as ``Specialized, Refined, Distinctive, and Innovative''. Our analysis focuses on the subset designated as ``Little Giant'' firms, comprising small businesses exhibiting strong performance and growth potential. These firms receive substantial policy support, generating structural discontinuities in debt financing cost that warrant explicit control.}

\autoref{tabExcludeSRDI} reports four specifications isolating the AI policy effect from SRDI confounds. Column (1) presents the benchmark estimate. Column (2) controls for batch-specific SRDI policy effects by including interaction terms between each designation cohort (2019 -- 2024) and the post-policy dummy, thereby accounting for dynamic certification shocks across years. Column (3) excludes firms designated in the 2021 cohort, which directly overlaps with the AI policy announcement, ensuring that the treatment effect is not driven by contemporaneous certification status. Column (4) implements a Difference-in-Difference-in-Differences (DDD) specification that includes the AI washing indicator, the 2021 SRDI policy effect, and their interaction, thereby isolating whether the AI policy effect differs systematically within the 2021-certified subsample.

\begin{table}[htbp]
	\centering
	\caption{\centering Excluding ``SRDI Little Giant'' Recognition}
    \setlength{\tabcolsep}{2pt}
    \renewcommand{\arraystretch}{1.15}
    \footnotesize
	\begin{tabular}{l*4{>{\centering\arraybackslash}p{3.33cm}}}
		\toprule
		 & (1) & (2) & (3) & (4) \\
		 & Debt FC & Debt FC & Debt FC & Debt FC \\
		\midrule
		AI Washing & 0.125*** & 0.125*** & 0.129*** & 0.128*** \\
		& (0.036) & (0.036) & (0.036) & (0.036) \\
		AI Washing\_SRDI &  &  &  & -0.144 \\
		&  &  &  & (0.188) \\
		SRDI\_Post\_2019 &  & 0.172 &  &  \\
		&  & (0.165) &  &  \\
		SRDI\_Post\_2020 &  & 0.161* &  &  \\
		&  & (0.097) &  &  \\
		SRDI\_Post\_2021 &  & 0.197** &  & 0.240* \\
		&  & (0.098) &  & (0.128) \\
		SRDI\_Post\_2022 &  & 0.058 &  &  \\
		&  & (0.069) &  &  \\
		SRDI\_Post\_2023 &  & -0.012 &  &  \\
		&  & (0.145) &  &  \\
		SRDI\_Post\_2024 &  & 0.036 &  &  \\
        \midrule
		Controls & Yes & Yes & Yes & Yes \\
		Firm FE & Yes & Yes & Yes & Yes \\
		Year FE & Yes & Yes & Yes & Yes \\
		Year\_Ind FE & Yes & Yes & Yes & Yes \\
		Year\_Prov FE & Yes & Yes & Yes & Yes \\
		$N$ & 34924 & 34924 & 34199 & 34924 \\
		Adj.\;$R^{2}$ & 0.498 & 0.498 & 0.497 & 0.498 \\
		\bottomrule
        \addlinespace[0.1em]
        \multicolumn{5}{l}{\footnotesize $^{*}$ $p<0.1$, $^{**}$ $p<0.05$, $^{***}$ $p<0.01$} \\
        \multicolumn{5}{l}{\footnotesize Note: Standard errors clustered at firm level in parentheses.} \\
	\end{tabular}\label{tabExcludeSRDI}
\end{table}

The estimated coefficient on \textit{AI Washing} varies minimally between 0.125 and 0.129 across all four specifications, remaining precisely estimated at the 1\% level. This stability indicates that the relative increase in debt financing cost for AI washing firms is not an artifact of preferential credit certification. The insensitivity of the treatment effect to both dynamic cohort controls and the exclusion of overlapping 2021 designees confirms that the documented market discipline operates through the strategic priority channel rather than through firm-specific subsidy programs.

\subsubsection{Excluding Real Estate ``Three Red Lines'' Policy}

Second, we address the real estate ``Three Red Lines'' policy introduced in 2020, which imposed stringent balance sheet constraints on property developers and generated sector-specific credit tightening. Because this intervention altered the aggregate credit supply landscape during our treatment window, it may confound inferences regarding the AI policy effect if debt repricing in the real estate sector spilled over into broader credit markets.

\autoref{tabRealEstatePolicy} reports four specifications paralleling the SRDI analysis. Column (1) presents the benchmark estimate. Column (2) controls for the policy impact using the \textit{RE\_Post} indicator, which interacts a real estate sector dummy with the post-policy period. The coefficient on \textit{RE\_Post} is positive and significant at the 1\% level, confirming independent borrowing cost increases for affected firms. Column (3) excludes real estate firms entirely from the sample. Column (4) implements a DDD specification including the AI washing indicator, the real estate policy effect, and their interaction.

\begin{table}[htbp]
	\centering
	\caption{\centering Excluding Real Estate ``Three Red Lines'' Policy}
    \setlength{\tabcolsep}{2pt}
    \renewcommand{\arraystretch}{1.15}
    \footnotesize
	\begin{tabular}{l*4{>{\centering\arraybackslash}p{3.4cm}}}
		\toprule
		 & (1) & (2) & (3) & (4) \\
		 & Debt FC & Debt FC & Debt FC & Debt FC \\
		\midrule
		AI Washing & 0.127*** & 0.130*** & 0.124*** & 0.125*** \\
		& (0.035) & (0.035) & (0.034) & (0.034) \\
		AI Washing\_RE &  &  &  & 0.179 \\
		&  &  &  & (0.362) \\
		RE\_Post &  & 0.521*** &  & 0.458*** \\
		&  & (0.139) &  & (0.121) \\
		\midrule
		Controls & Yes & Yes & Yes & Yes \\
		Firm FE & Yes & Yes & Yes & Yes \\
		Year FE & Yes & Yes & Yes & Yes \\
		Year\_Prov FE & Yes & Yes & Yes & Yes \\
		$N$ & 34924 & 34924 & 33782 & 34924 \\
		Adj.\;$R^{2}$ & 0.492 & 0.493 & 0.504 & 0.493 \\
		\bottomrule
        \addlinespace[0.1em]
        \multicolumn{5}{l}{\footnotesize $^{*}$ $p<0.1$, $^{**}$ $p<0.05$, $^{***}$ $p<0.01$} \\
        \multicolumn{5}{l}{\footnotesize Note: Standard errors clustered at firm level in parentheses.} \\
	\end{tabular}\label{tabRealEstatePolicy}
\end{table}

The point estimate on \textit{AI Washing} remains tightly bounded between 0.124 and 0.130 throughout, retaining significance at the 1\% level even after purging real estate sector effects. The absence of meaningful attenuation when controlling for or excluding property developers demonstrates that sector-specific deleveraging shocks do not explain the financing cost penalty for AI washing. Furthermore, the statistically insignificant DDD interaction confirms that the AI policy treatment is orthogonal to real estate exposure, validating that the identified market discipline is not contaminated by concurrent credit tightening in the property sector.

\subsubsection{Excluding Green Credit Policy}

Finally, we examine the green credit policy, which differentially imposes stricter financing conditions on high-polluting firms. If creditors conflated environmental noncompliance with AI washing during the post-policy period, the estimated treatment effect might capture green financing repricing rather than strategic priority-driven screening.

\autoref{tabGreenCreditPolicy} reports four specifications paralleling the prior analyses. Column (1) presents the benchmark estimate. Column (2) controls for the policy impact using the \textit{Pollute\_Post} indicator, which interacts a high-pollution sector dummy with the post-policy period. The coefficient on \textit{Pollute\_Post} is negative and significant at the 1\% level, indicating that polluting firms face tighter financing conditions following the policy. Column (3) excludes high-polluting firms entirely from the sample. Column (4) implements a DDD specification including the AI washing indicator, the green credit policy effect, and their interaction.

\begin{table}[htbp]
	\centering
	\caption{\centering Excluding Green Credit Policy}
    \setlength{\tabcolsep}{2pt}
    \renewcommand{\arraystretch}{1.15}
    \footnotesize
	\begin{tabular}{l*4{>{\centering\arraybackslash}p{3.4cm}}}
		\toprule
		 & (1) & (2) & (3) & (4) \\
		 & Debt FC & Debt FC & Debt FC & Debt FC \\
		\midrule
		AI Washing & 0.125*** & 0.106*** & 0.117*** & 0.100** \\
		& (0.036) & (0.036) & (0.040) & (0.040) \\
		AI Washing\_GC &  &  &  & 0.033 \\
		&  &  &  & (0.091) \\
		GC\_Post &  & -0.270*** &  & -0.278*** \\
		&  & (0.051) &  & (0.058) \\
		\midrule
		Controls & Yes & Yes & Yes & Yes \\
		Firm FE & Yes & Yes & Yes & Yes \\
		Year FE & Yes & Yes & Yes & Yes \\
		Year\_Ind FE & Yes & Yes & Yes & Yes \\
		Year\_Prov FE & Yes & Yes & Yes & Yes \\
		$N$ & 34924 & 34924 & 27505 & 34924 \\
		Adj.\;$R^{2}$ & 0.498 & 0.499 & 0.478 & 0.499 \\
		\bottomrule
        \addlinespace[0.1em]
        \multicolumn{5}{l}{\footnotesize $^{*}$ $p<0.1$, $^{**}$ $p<0.05$, $^{***}$ $p<0.01$} \\
        \multicolumn{5}{l}{\footnotesize Note: Standard errors clustered at firm level in parentheses.} \\
	\end{tabular}\label{tabGreenCreditPolicy}
\end{table}

The coefficient on \textit{AI Washing} retains positive significance across all four columns, ranging from 0.100 to 0.125. While Column (2) exhibits modest attenuation relative to the benchmark, the coefficient remains significant at the 1\% level, and the full-sample DDD specification in Column (4) confirms that the AI policy effect does not interact with pollution status. The stability of the treatment effect after purging environmental regulatory confounds indicates that the documented relative increase in debt financing cost is not an epiphenomenon of green credit tightening but rather reflects the distinctive information shock generated by the strategic elevation of artificial intelligence.

\section{Further Explorations}\label{Further Explorations}

\subsection{Cross-Equation Joint Tests via Seemingly Unrelated Regressions}

We employ seemingly unrelated regressions \citep{zellner1962efficient} to jointly estimate the multidimensional adjustments mandated by \textbf{H2}. The system comprises four standardized dependent variables, relative debt financing cost, debt financing flows, AI narrative intensity, and AI patent stock. Single-equation estimation would yield inefficient estimates and potentially biased inference if unobserved shocks to debt contracting, disclosure strategies, and innovation investment are mutually correlated. The Breusch-Pagan test reported in \autoref{tab:res_corr} strongly rejects error orthogonality, confirming that disturbances across equations share common components and validating the SUR approach relative to equation-by-equation OLS.

\begin{table}[htbp]
	\centering
	\caption{\centering Joint Estimation via Seemingly Unrelated Regressions}
	\setlength{\tabcolsep}{2pt}
	\renewcommand{\arraystretch}{1.15}
    \footnotesize
	\begin{tabular}{l*4{>{\centering\arraybackslash}p{3.5cm}}}
		\toprule
		 & Debt Cost & Debt Flow & AI Words & AI Patents \\
		\midrule
		AI Washing & 0.018*** & -0.049*** & -0.044*** & 0.051*** \\
		& (0.006) & (0.007) & (0.004) & (0.004) \\
		\midrule
		Controls & Yes & Yes & Yes & Yes \\
		Firm FE & Yes & Yes & Yes & Yes \\
		Year FE & Yes & Yes & Yes & Yes \\
		Year\_Ind FE & Yes & Yes & Yes & Yes \\
		Year\_Prov FE & Yes & Yes & Yes & Yes \\
        $N$ & 34,924 & 34,924 & 34,924 & 34,924 \\
		\bottomrule
        \addlinespace[0.1em]
        \multicolumn{5}{l}{\footnotesize $^{*}$ $p<0.1$, $^{**}$ $p<0.05$, $^{***}$ $p<0.01$.} \\
        \multicolumn{5}{l}{\footnotesize Notes: (1) Dependent variables standardized to Z-scores.} \\
        \multicolumn{5}{l}{\footnotesize (2) Firm FE is controlled via within-demeaning.} \\
        \multicolumn{5}{l}{\footnotesize (3) Robust standard errors in parentheses.} \\
	\end{tabular}\label{sur}
\end{table}

\autoref{sur} presents the joint estimation results. The coefficient on AI washing is positive and significant in the debt financing cost equation, negative and significant in the debt financing flows equation, negative and significant in the AI narrative intensity equation, and positive and significant in the AI patenting equation. This quadruple pattern corroborates all dimensions of \textbf{H2}. AI washing firms simultaneously experience elevated relative financing costs alongside restricted debt financing flows while reducing symbolic disclosure intensity and increasing substantive patenting activity.

The standardized coefficients reveal an asymmetric adjustment pattern. The magnitude of patent accumulation exceeds that of narrative retrenchment, and the quantity constraint on debt financing exceeds the marginal cost increase. This suggests that financing constraints dominate pricing adjustments in compelling corporate resource reallocation. Rather than uniformly adjusting across margins, firms prioritize substantive capability demonstration over symbolic disclosure reduction, facing more pronounced credit rationing than interest rate penalties. 

\begin{table}[htbp]
	\centering
	\caption{\centering Residual Correlation Matrix from SUR Estimation}
	\setlength{\tabcolsep}{2pt}
	\renewcommand{\arraystretch}{1.25}
    \footnotesize
	\begin{tabular}{l*{4}{>{\centering\arraybackslash}p{3.5cm}}}
		\toprule
		& Debt Cost & Debt Flow & AI Words & AI Patents \\
		\midrule
		Debt Cost & 1.0000 & & & \\
		Debt Flow & 0.3025*** & 1.0000 & & \\
		AI Words & -0.0058 & 0.0020 & 1.0000 & \\
		AI Patents & 0.0092 & 0.0176 & 0.0257 & 1.0000 \\
		\midrule
		\multicolumn{5}{l}{Breusch-Pagan test of independence, $\chi^2(6) = 3234.9$***} \\
		\bottomrule
        \addlinespace[0.1em]
        \multicolumn{5}{l}{\footnotesize $^{*}$ $p<0.1$, $^{**}$ $p<0.05$, $^{***}$ $p<0.01$.} \\
        \multicolumn{5}{l}{\footnotesize Note: Breusch-Pagan $\chi^2$ rejects error independence, justifying SUR.} \\
	\end{tabular}\label{tab:res_corr}
\end{table}

\autoref{sur_tests} reports formal cross-equation hypothesis tests. Panel A confirms that each individual coefficient differs significantly from zero, with the joint test of all four coefficients being zero strongly rejected. This validates that the policy shock induces a statistically coherent multidimensional adjustment. Panel B tests the equality constraint that absolute adjustment magnitudes are identical across dimensions. Rejection of this constraint indicates that firms do not adjust uniformly across margins. They instead exhibit selective investment behavior under financing constraints, prioritizing patent accumulation over narrative retrenchment and facing more severe quantity rationing than marginal cost increases. The dominance of quantity rationing over price penalties aligns with \citet{stiglitz1981credit} where adverse selection prevents market clearing via interest rates alone, compelling creditors to simultaneously restrict credit availability. This asymmetric adjustment supports the theoretical prediction that credit market frictions drive efficient technological pivots from symbolic disclosure toward verifiable innovation.

\begin{table}[htbp]
	\centering
	\caption{\centering Cross-Equation Hypothesis Tests}
	\setlength{\tabcolsep}{3pt}
	\renewcommand{\arraystretch}{1.25}
    \footnotesize
	\begin{tabular}{l*3{>{\centering\arraybackslash}p{3.85cm}}}
		\toprule
		\textbf{Test} & \textbf{$\chi^2$ Statistic} & \textbf{df} & \textbf{$p$-value} \\
		\midrule
        \addlinespace
		\multicolumn{3}{l}{\textit{\textbf{A. Individual and Joint Tests}}} \\
		$\beta_{\text{cost}} =0 $ & 10.07 & 1 & 0.006*** \\
		$\beta_{\text{flow}} =0 $ & 46.31 & 1 & 0.000*** \\
		$\beta_{\text{word}} =0$ & 149.85 & 1 & 0.000*** \\
		$\beta_{\text{patent}} =0$ & 184.79 & 1 & 0.000*** \\
		$\beta_{\text{cost}} = \beta_{\text{flow}} = \beta_{\text{word}} =\beta_{\text{patent}} =0$ & 387.85 & 4 & 0.000*** \\
		\midrule
        \addlinespace
		\multicolumn{3}{l}{\textit{\textbf{B. Equality Constraints Test}}} \\
		$\beta_{\text{cost}}=\beta_{\text{flow}}=-\beta_{\text{word}}=\beta_{\text{patent}}$ & 198.21 & 3 & 0.000*** \\
		\bottomrule
        \addlinespace[0.1em]
        \multicolumn{3}{l}{\footnotesize $^{*}$ $p<0.1$, $^{**}$ $p<0.05$, $^{***}$ $p<0.01$.} \\
        \multicolumn{3}{l}{\footnotesize Note, Individual tests use Bonferroni-adjusted $p$-values.} \\
	\end{tabular}\label{sur_tests}
\end{table}

\subsection{Mechanism Analysis}

\begin{table}[htbp]
	\centering
	\caption{\centering Moderating Effect Analysis}
    \setlength{\tabcolsep}{2pt}
    \footnotesize
    \renewcommand{\arraystretch}{1.15}
	\begin{tabular}{l*6{>{\centering\arraybackslash}p{2cm}}}
		\toprule
      & \multicolumn{2}{c}{Attention Reallocation} & \multicolumn{2}{c}{Relationship (``Guanxi'')} & \multicolumn{2}{c}{Regulatory Channel} \\
      \cmidrule(r){2-3} \cmidrule(r){4-5} \cmidrule(r){6-7}
		 & Debt FC & Debt FC  & Debt FC & Debt FC & Violation & Inquiry \\
		\midrule
		AI Washing & 0.082** & 0.076* & 0.222*** & 0.313*** & 0.121*** & 0.132*** \\
		& (0.041) & (0.045) & (0.064) & (0.095) & (0.035) & (0.036) \\
		AI Washing\_Mshare & 0.364*** &  &  &  &  &  \\
		& (0.128) &  &  &  &  &  \\
		AI Washing\_ATT &  & 0.037** &  &  &  &  \\
		&  & (0.019) &  &  &  &  \\
		AI Washing\_SC &  &  & -0.307* &  &  &  \\
		&  &  & (0.171) &  &  &  \\
		AI Washing\_Bank &  &  &  & -0.198** &  &  \\
		&  &  &  & (0.095) &  &  \\
		AI Washing\_Vio &  &  &  &  & 0.044 &  \\
		&  &  &  &  & (0.081) &  \\
		AI Washing\_Inq &  &  &  &  &  & -0.097 \\
		&  &  &  &  &  & (0.089) \\
		Mshare & -0.267** &  &  &  &  &  \\
		& (0.120) &  &  &  &  &  \\
		ATT &  & -0.071*** &  &  &  &  \\
		&  & (0.013) &  &  &  &  \\
		SC &  &  & -0.433*** &  &  &  \\
		&  &  & (0.137) &  &  &  \\
		Bank &  &  &  & 0.057 &  &  \\
		&  &  &  & (0.070) &  &  \\
        Violation &  &  &  &  & 0.112*** &  \\
		&  &  &  &  & (0.031) &  \\
		Inquiry &  &  &  &  &  & 0.190*** \\
		&  &  &  &  &  & (0.042) \\
        \midrule
        Controls & Yes & Yes & Yes & Yes & Yes & Yes \\
		Firm FE & Yes & Yes & Yes & Yes & Yes & Yes \\
		Year FE & Yes & Yes & Yes & Yes & Yes & Yes \\
		Year\_Ind FE & Yes & Yes & Yes & Yes & Yes & Yes \\
		Year\_Prov FE & Yes & Yes & Yes & Yes & Yes & Yes \\
		$N$ & 34924 & 34924 & 34924 & 34924 & 34924 & 34924 \\
		Adj.\;$R^{2}$ & 0.498 & 0.499 & 0.499 & 0.498 & 0.499 & 0.499 \\
		\bottomrule
        \addlinespace[0.1em]
        \multicolumn{3}{l}{\footnotesize $^{*}$ $p<0.1$, $^{**}$ $p<0.05$, $^{***}$ $p<0.01$} \\
        \multicolumn{3}{l}{\footnotesize Note: Standard errors clustered at firm level in parentheses.} \\
	\end{tabular}\label{tabModerating}
\end{table}

We examine heterogeneity in relative financing cost adjustments across signal credibility and relational assurance mechanisms. \autoref{tabModerating} reports interaction specifications testing \textbf{H3} and \textbf{H4}.

Columns (1) and (2) examine the attention reallocation mechanism from complementary angles. Column (1) presents the internal governance signal. The main effect of management shareholding is significantly negative at minus 0.267, consistent with the classical agency theory prediction that insider equity stakes align managerial incentives with firm value and reduce financing cost \citep{JENSEN1976305}. However, the interaction coefficient \textit{AI Washing} $\times$ \textit{Mshare} is 0.364 and significant at the 1\% level, indicating that the financing cost penalty for AI washing is substantially amplified among firms with concentrated management shareholding. This sign reversal reveals that under heightened policy-induced scrutiny, high management shareholding transforms from a governance remedy into a liability signal. Managers with substantial equity positions possess both the technical sophistication to understand AI capabilities and the strategic incentive to exploit information asymmetry. Creditors therefore interpret the coexistence of AI washing and concentrated insider stakes as informed strategic misconduct rather than uninformed disclosure error, elevating the inferred risk of intentional narrative manipulation. Column (2) presents the external information environment. The main effect of analyst attention is significantly negative at minus 0.071, reflecting the standard information intermediary function through which analyst attention level reduces information asymmetry and lowers financing costs \citep{cheng2008analyst}. Yet the interaction coefficient \textit{AI Washing} $\times$ \textit{ATT} is 0.037 and significant at the 5\% level, indicating that elevated analyst attention amplifies the financing cost penalty for AI washing firms. When analysts concentrate scrutiny on specific firms, creditors face heightened reputational pressure to detect and penalize decoupled disclosures, as screening failures in high-visibility environments carry greater external audience costs. The policy-induced attention crowding-in thus operates through both internal governance channels and external information amplification, jointly supporting \textbf{H3}.

Columns (5) and (6) examine whether contemporaneous regulatory enforcement drives the financing cost penalty. The interaction coefficients \textit{AI Washing} $\times$ \textit{Violation} and  \textit{AI Washing} $\times$ \textit{Inquiry} are both statistically insignificant. These null interactions indicate that the relative increase in debt financing costs for AI washing firms is not amplified by concurrent regulatory penalties or inquiries. While the main effects of \textit{Violation} and \textit{Inquiry} are individually positive and significant, confirming that regulatory actions independently increase financing costs, they do not interact with AI washing status to produce differential pricing. This finding is critical because it rules out the alternative interpretation that the documented treatment effect reflects direct regulatory punishment rather than market-based screening. The absence of a significant interaction demonstrates that the policy shock operates through information environment restructuring (\textbf{H3}) rather than through contemporaneous enforcement channels.

Columns (3) and (4) examine the relational governance mechanism from complementary angles. Column (3) presents the supply chain dimension. The main effect of supply chain concentration is significantly negative at minus 0.433, confirming that dense buyer-supplier relationships independently reduce financing costs by generating relational capital that functions as implicit collateral. The interaction coefficient \textit{AI Washing} $\times$ \textit{SC} is minus 0.307 and significant at the 10\% level, indicating that supply chain concentration further attenuates the financing cost penalty for AI washing firms. The joint significance of both terms demonstrates that supply chain relationships constitute an actively cultivated governance mechanism rather than a passive industrial structure. Firms embedded in concentrated transactional networks have repeatedly demonstrated their technological substance to upstream and downstream partners, enabling creditors to rely on verified commercial interactions as alternative assurance channels that substitute for direct AI capability screening. Column (4) presents the banking proximity dimension. The main effect of bank proximity is statistically indistinguishable from zero at 0.057, indicating that mere geographic proximity to banking institutions does not independently alter financing costs. This null main effect is critical because it rules out the alternative interpretation that regional financial development drives the observed heterogeneity. The interaction coefficient \textit{AI Washing} $\times$ \textit{Bank} is minus 0.198 and significant at the 5\% level, revealing that banking proximity attenuates the financing cost penalty exclusively in the post-policy period. This conditional effect confirms that the operative mechanism is not macro-level financial depth but micro-level soft information production. Localized bank presence facilitates repeated interactions and community-based monitoring that generate non-verifiable intelligence about firm capabilities, enabling creditors to distinguish genuine AI adopters from symbolic impostors without relying solely on formal disclosure. Taken together, these results demonstrate that relational governance buffers the financing cost penalty through both internal supply chain trust and external banking network embeddedness, jointly supporting \textbf{H4}.

\subsection{Heterogeneity Analysis}

We examine whether the debt financing cost penalty varies with persistent firm characteristics measured during the pre-policy period. The empirical $p$-values reported in \autoref{tabHeterogeneity} test the significance of coefficient differences across subsamples defined by time-averaged pre-treatment characteristics.

\begin{table}[htbp]
    \centering
    \caption{\centering Heterogeneity Analysis}
    \vspace{0.5em}
    \setlength{\tabcolsep}{2pt}
    \footnotesize
    \renewcommand{\arraystretch}{1.15}
    \begin{tabular}{l*{8}{>{\centering\arraybackslash}p{1.62cm}}}
		\toprule
        & \multicolumn{4}{c}{Deep-level Signals} & \multicolumn{4}{c}{Relationship (Guanxi)} \\
        \cmidrule(r){2-5} \cmidrule(l){6-9}
		& \multicolumn{2}{c}{Patent Non-Self Citations} & \multicolumn{2}{c}{Disruptive Innovation} & \multicolumn{2}{c}{With Merchant Guilds} & \multicolumn{2}{c}{With Government} \\
		\cmidrule(r){2-3} \cmidrule(l){4-5} \cmidrule(l){6-7} \cmidrule(l){8-9}
		& Low & High & Low & High & Low & High & Low & High \\
        \midrule
		AI Washing & 0.202*** & 0.043 & 0.215*** & 0.060 & 0.193*** & 0.070 & 0.241*** & -0.050 \\
		& (0.060) & (0.043) & (0.055) & (0.048) & (0.053) & (0.050) & (0.043) & (0.061) \\
		Size & -0.061 & -0.037 & -0.040 & -0.081 & 0.026 & -0.089 & -0.107 & 0.061 \\
		& (0.090) & (0.067) & (0.100) & (0.069) & (0.083) & (0.071) & (0.079) & (0.061) \\
		Lev & 1.295*** & 1.084*** & 1.051*** & 1.323*** & 1.267*** & 1.060*** & 1.443*** & 0.760*** \\
		& (0.214) & (0.171) & (0.192) & (0.191) & (0.196) & (0.193) & (0.165) & (0.234) \\
		ROA & -1.073*** & -1.125*** & -1.014*** & -1.152*** & -0.959*** & -1.251*** & -0.979*** & -1.402*** \\
		& (0.257) & (0.363) & (0.203) & (0.421) & (0.226) & (0.407) & (0.303) & (0.296) \\
		Liquid & 0.026 & -0.014* & 0.020 & -0.013 & 0.004 & 0.014 & 0.019 & -0.022* \\
		& (0.018) & (0.007) & (0.017) & (0.009) & (0.010) & (0.019) & (0.014) & (0.013) \\
		Top5 & -0.415 & -1.146*** & -0.873*** & -0.599** & -1.006*** & -0.601** & -0.701*** & -1.050*** \\
		& (0.253) & (0.238) & (0.243) & (0.257) & (0.233) & (0.242) & (0.204) & (0.291) \\
		TobinQ & -0.002 & -0.028 & 0.005 & -0.021 & 0.006 & -0.022* & -0.017 & 0.015 \\
		& (0.016) & (0.027) & (0.019) & (0.016) & (0.024) & (0.013) & (0.012) & (0.040) \\
		ListAge & 0.204*** & 0.211*** & 0.109* & 0.384*** & 0.207*** & 0.212*** & 0.209*** & 0.149** \\
		& (0.056) & (0.063) & (0.062) & (0.092) & (0.055) & (0.059) & (0.050) & (0.076) \\
		\_cons & 2.229 & 2.287* & 2.205 & 2.469* & 0.655 & 3.072** & 3.305** & 0.371 \\
		& (1.890) & (1.379) & (2.131) & (1.440) & (1.773) & (1.465) & (1.649) & (1.306) \\
        \midrule
		Firm FE & Yes & Yes & Yes & Yes & Yes & Yes & Yes & Yes \\
		Year FE & Yes & Yes & Yes & Yes & Yes & Yes & Yes & Yes \\
		Year\_Ind FE & Yes & Yes & Yes & Yes & Yes & Yes & Yes & Yes \\
		Year\_Prov FE & Yes & Yes & Yes & Yes & Yes & Yes & Yes & Yes \\
		$N$ & 17463 & 17449 & 17529 & 17390 & 17459 & 17443 & 22564 & 12358 \\
		Adj.\;$R^{2}$ & 0.510 & 0.494 & 0.512 & 0.487 & 0.565 & 0.437 & 0.464 & 0.571 \\
        \cmidrule(r){2-3} \cmidrule(l){4-5} \cmidrule(l){6-7} \cmidrule(l){8-9}
        Empirical $p$-value & \multicolumn{2}{c}{0.002***} & \multicolumn{2}{c}{0.004***} & \multicolumn{2}{c}{0.008***} & \multicolumn{2}{c}{0.000***} \\
        \bottomrule
        \addlinespace[0.1em]
        \multicolumn{9}{l}{\footnotesize * $p<0.1$, ** $p<0.05$, *** $p<0.01$} \\
        \multicolumn{9}{l}{\footnotesize Note: Standard errors clustered at firm level in parentheses.} \\
	\end{tabular}
    \label{tabHeterogeneity}
\end{table}

The deep-level signal dimension extends the attention reallocation logic underlying \textbf{H3}. While management shareholding and analyst attention level capture the intensity of creditor scrutiny, the availability of verifiable capability signals determines whether that scrutiny translates into differential financing costs. Firms with low external patent citations face a severe debt financing cost penalty of 0.202 when engaging in AI washing, whereas firms with high external knowledge absorption exhibit no significant cost adjustment with a point estimate of 0.043. Similarly, firms lacking disruptive innovation face a penalty of 0.215, while firms with breakthrough technological orientation show no significant adjustment with a point estimate of 0.060. This pattern indicates that the attention crowding-in mechanism documented in \textbf{H3} imposes constraints primarily when deep-level quality indicators are absent. When firms possess external knowledge linkages or market-disrupting innovation, creditors utilize these alternative verification channels to assess AI capabilities, rendering the policy-induced scrutiny less consequential for financing costs.

The relational governance dimension extends the buffering logic underlying \textbf{H4}. While supply chain concentration and bank proximity capture the moderating effect of relational ties on the financing cost penalty, the heterogeneity analysis reveals how these ties shape the cross-sectional distribution of market discipline. Firms lacking strong merchant guild affiliations face a substantial penalty of 0.193, whereas firms embedded in dense guild networks experience an attenuated and insignificant penalty of 0.070. This indicates that historically concentrated merchant communities provide reputational insurance that substitutes for formal credit risk assessment, paralleling the supply chain trust mechanism in \textbf{H4}. The government relationship partition reveals an even starker contrast: firms lacking political connections face a coefficient of 0.241, while connected firms exhibit an insignificant negative coefficient of minus 0.050. This suggests that state-affiliated firms enjoy implicit guarantees or regulatory forbearance that insulate them from market discipline regarding disclosure authenticity, extending the relational buffering logic of \textbf{H4} to the political domain.

These findings reveal a dual characteristic of the policy-driven credit market adjustment. Creditors demonstrate sophisticated capacity to penetrate surface-level narrative disclosures and rely on deep-level capability signals accumulated prior to the policy shock, penalizing firms with weak technological substance. Simultaneously, the financing market exhibits respect for the distinctively Chinese relational governance structures, which function as informal insurance mechanisms. This coexistence of hard-information scrutiny and relational governance highlights an institutional complementarity that distinguishes China's emerging credit market from Western contexts where formal contracting and hard information dominate.

\section{Discussion and Conclusion}\label{Conclusion}

\subsection{Discussion}

The validity of our residual based measure rests on whether it captures strategic decoupling rather than random innovation volatility. The robustness patches reported in section \ref{Empirical Results} and Appendix \ref{ApdxD} strongly support the former. The strict treatment definition requiring strictly positive residuals across all prepolicy years and the extreme upper tercile subsample yield economically comparable or larger coefficients than the baseline. Excluding the year 2020 further strengthens the effect indicating that the documented market discipline targets habitual strategic behavior rather than pandemic induced noise. Moreover the residuals predict future narrative inflation and patent declines and correlate with lower prepolicy financing costs. These patterns collectively establish that the residual identifies firms that systematically inflate AI disclosure relative to substantive patenting capability.

Our findings illuminate a soft governance mechanism operating through information coordination. By elevating AI to strategic national priority the 14th Five Year Plan increases the marginal return to creditor screening in this domain without relying on explicit mandates or punitive enforcement. The statistically insignificant interactions between AI washing and contemporaneous regulatory violations or inquiries rule out direct regulatory punishment as the driving force. Instead creditors autonomously reallocate attention toward AI capability verification thereby imposing market discipline through endogenous screening intensification.

The seemingly unrelated regressions reveal multidimensional corporate adjustments. AI washing firms simultaneously face higher relative financing costs tighter debt financing flows reduced AI narrative intensity and increased AI patenting. Cross equation tests reject equal adjustment magnitudes across margins indicating that quantity rationing dominates price penalties. This asymmetric pattern aligns with credit rationing theory and suggests that financing constraints compel firms to prioritize verifiable capability demonstration over symbolic disclosure reduction.

Heterogeneity analysis uncovers a dual institutional logic distinctive to emerging markets. The attention crowding in mechanism concentrates penalties among firms lacking deep level signals such as external patent citations and disruptive innovation capacity. Conversely relational governance structures attenuate this effect. Supply chain concentration generates accumulated trust that substitutes for direct technological verification while geographic proximity to banks facilitates soft information production through community based monitoring. Merchant guild affiliation and government connections further buffer the financing cost penalty by providing reputational insurance and implicit guarantees. This coexistence of hard information scrutiny and relational buffering demonstrates that capital allocation efficiency in China depends not on replicating western mode but on the joint operation of policy induced attention shocks and indigenous network based governance.

\subsection{Conclusion}

This study examines how macro level strategic policy shocks activate micro level market discipline in corporate debt markets. Leveraging China's 14th Five Year Plan as a quasi natural experiment we document that firms with historically decoupled AI activities experience a relative increase in debt financing cost of 12.5 to 19.3 basis points post shock. These firms simultaneously reduce symbolic AI disclosure increase substantive AI patenting and face tightened credit rationing.

Our contributions are fourfold. First we extend the residual based approach from the earnings management literature to measure AI decoupling and validate its external fraudulence through subsidy extraction regulatory violation concealment and evasion of contemporaneous oversight. Second we establish that strategic priority signals function as information coordination devices capable of activating market discipline absent direct regulatory enforcement. Third we isolate the transmission channel and confirm that information driven creditor screening rather than contemporaneous regulatory punishment drives the documented effects. Fourth we challenge the conventional view that emerging market debt markets cannot assess complex technological claims by revealing a dual institutional complementarity between hard information scrutiny and relational governance.

Several limitations remain. The residual based measure is an inferred proxy for managerial intent and future research could exploit hand collected evidence or regulatory archives to refine measurement. Whether comparable market discipline arises in other emerging economies or technological domains such as low-altitude technology or biotechnology warrants investigation. Finally the long run dynamic consequences of AI washing discipline for firm survival and innovation trajectories deserve continued attention as postpolicy data accumulate.

\newpage
\appendix

\section{Description and Definition of Variables}\label{ApdxA}
The detailed description and definition of the variables related to this study are presented in \autoref{tabDefVar}.
\begin{table}[htbp]
    \centering
    \caption{\centering Description and Definition of Variables}
    \label{tabDefVar}
    \vspace{0.5em}
    \setlength{\tabcolsep}{2pt}
    \renewcommand{\arraystretch}{1.1}
    \footnotesize
    \begin{tabular}{l*1{>{\arraybackslash}p{12.5cm}}}
    \toprule
         \textbf{Variable} &  \textbf{Definition} \\
    \midrule
         \textit{Debt FC} & Ratio of total debt-related expenditures to end-of-period total liabilities, expressed in \%. \\
         \textit{AI Word} & $\ln$ of one plus the frequency of AI-related keywords in annual reports. \\
         \textit{AI Patent} & $\ln$ of one plus the count of AI invention patents obtained both independently and jointly. \\
         \textit{Size} & $\ln$ of total assets. \\
         \textit{Lev} & Total liabilities over total assets. \\
         \textit{ROA} & Net profit over total assets. \\
         \textit{Liquid} & Ratio of current assets to current liabilities. \\
         \textit{Top5} & Combined shareholding percentage of the top five shareholders. \\
         \textit{TobinQ} & Market value of assets over the book value of assets. \\
         \textit{ListAge} & $\ln$ of one plus the number of years since the firm's IPO.\\
        \textit{IT\_ratio} & Ratio of executives with IT backgrounds among the total executive team. \\
        \textit{no\_entry} & Dummy variable equal to one if the firm was listed in or after 2021. \\
        \textit{Debt Flow} & Ratio of newly incurred liabilities to total assets at fiscal year-end, expressed in \%.
        \\
        \textit{Innov\_Sub} & Dummy variable equal to one if the firm receives innovation subsidy from the government. \\
        \textit{Violation} & Dummy variable equal to one if the firm is punished. \\
        \textit{Inquiry} & Dummy variable equal to one if the firm is inquiried by regulators. \\
        \textit{Mshare} & Management’s shareholding. \\
        \textit{ATT} & $\ln$ of one plus the number of analysts who focus on a company. \\
        \textit{SC} & Supply chain concentration level, expressed as mean of the sum of purchase from the top five suppliers and sales to the top five customers. \\
        \textit{Bank} & Dummy variable equal to one of there is any type of bank within 3km distance of the company's registered address. \\
        \textit{Patent Non-Self Citations} & Mean of external citations received by the firm's patents across years, excluding self-citations. \\
        \textit{Disruptive Innovation} & $\ln$ of one plus the count of patents classified under International Patent Classification (IPC) codes that were newly emerging during the 5-year window.\\
        \textit{Merchant Guilds Relationship} & Minimum geographic distance from firm's registered location to historical merchant guild centers.\\
        \textit{Government Relationship} & Dummy variable equal to one if either chairperson or CEO is a current or former government official.\\
    \bottomrule
    \end{tabular}
\end{table}

\section{Persistence of AI Washing Behavior}\label{ApdxC}

A potential concern with our time-invariant treatment assignment is that annual fluctuations in residuals may misclassify firms with transitory deviations as habitual AI washers. We address this concern by examining the year-to-year persistence of decoupling behavior using transition matrices and rank correlation analysis.

\begin{table}[htbp]
	\centering
	\caption{\centering Transition Matrix of Annual AI Washing Residuals (Terciles)}
    \setlength{\tabcolsep}{8pt}
    \renewcommand{\arraystretch}{1.15}
    \footnotesize
	\begin{tabular}{l*{3}{>{\centering\arraybackslash}p{3.9cm}}}
		\toprule
		& \multicolumn{3}{c}{Year $t$ (Current)} \\
		\cmidrule(lr){2-4}
		Year $t-1$ (Lagged) & Low Tercile & Middle Tercile & High Tercile \\
		\midrule
		Low Tercile & \textbf{71.85\%} & 19.05\% & 9.10\% \\
		Middle Tercile & 19.67\% & \textbf{62.57\%} & 17.76\% \\
		High Tercile & 8.48\% & 18.38\% & \textbf{73.14\%} \\
		\midrule
		\multicolumn{4}{l}{Spearman's $\rho = 0.7167^{***}$ (N = 15,526)} \\
		\bottomrule
        \addlinespace[0.1em]
        \multicolumn{4}{l}{\footnotesize $^{*}$ $p<0.1$, $^{**}$ $p<0.05$, $^{***}$ $p<0.01$} \\
        \multicolumn{4}{l}{\footnotesize Note: This table reports row-standardized transition probabilities across terciles of annual} \\
        \multicolumn{4}{l}{\footnotesize residuals. Bold values indicate persistence rates (diagonal elements).} \\
	\end{tabular}\label{tabTransition}
\end{table}

\autoref{tabTransition} reports the transition matrix of tercile-ranked residuals. The diagonal elements indicate substantial persistence. 71.85\% of firms in the lowest tercile remain there in the subsequent year, with only 9.10\% transitioning to the highest tercile. Similarly, 73.14\% of high-tercile firms persist, with merely 8.48\% falling to the lowest tercile. The Spearman rank correlation between consecutive years is 0.7167 and significant at the 1\% level, substantially exceeding the six-year average of 0.579 reported in the main text. This evidence confirms that while individual-year residuals contain measurement noise, firms exhibit persistent AI washing behavior over time, validating our use of multi-year averages to define treatment status and mitigating concerns about misclassification due to transitory fluctuations.

\section{More Validation of AI Decoupling Residuals}\label{ApdxB}

We validate the residual-based measure of AI decoupling as a proxy for AI washing by examining its pre-policy correlation with debt financing cost and its predictive power for subsequent technological and disclosure outcomes.

\begin{table}[htbp]
	\centering
	\caption{\centering AI Decoupling's Deceptive Effect on Debt Financing Pre-policy}
    \setlength{\tabcolsep}{2pt}
    \renewcommand{\arraystretch}{1.15}
    \footnotesize
	\begin{tabular}{l*7{>{\centering\arraybackslash}p{2cm}}}
		\toprule
		 & (1) & (2) & (3) & (4) & (5) & (6) & (7) \\
		 & 2015 & 2016 & 2017 & 2018 & 2019 & 2020 & Pooled \\
		\midrule
		res& -0.154*** & -0.116*** & -0.113*** & -0.090*** & -0.100*** & -0.070*** & -0.047** \\
		& (0.038) & (0.026) & (0.022) & (0.025) & (0.036) & (0.020) & (0.021) \\
		\midrule
		Controls & Yes & Yes & Yes & Yes & Yes & Yes & Yes \\
		FE & Ind & Ind & Ind & Ind & Ind & Ind & Year, Year\_Ind, Year\_Prov \\
		$N$ & 2690 & 2896 & 3312 & 3381 & 3509 & 3835 & 19192 \\
		Adj.\;$R^{2}$ & 0.135 & 0.137 & 0.156 & 0.182 & 0.080 & 0.191 & 0.513 \\
		\bottomrule
        \addlinespace[0.1em]
        \multicolumn{8}{l}{\footnotesize $^{*}$ $p<0.1$, $^{**}$ $p<0.05$, $^{***}$ $p<0.01$} \\
        \multicolumn{8}{l}{\footnotesize Note: Standard errors clustered at firm level in parentheses.} \\
	\end{tabular}\label{tabPretreatResidual}
\end{table}

\autoref{tabPretreatResidual} reports the pre-policy correlation between the decoupling residual and debt financing cost. The significantly negative coefficients across all years from 2015 to 2020 indicate that firms with higher pre-policy decoupling residuals systematically maintained lower debt financing cost prior to the policy intervention. This evidence confirms the existence of a financing privilege for AI washing firms in the baseline period, establishing the counterfactual condition for the dissipation of financing advantages documented in the main analysis. The consistency of this negative relationship across the pre-treatment window further validates that the residual captures inherent information risk rather than transitory measurement noise.

\begin{table}[htbp]
	\centering
	\caption{\centering Predictive Validity of AI Washing Residuals for Future Outcomes}
    \setlength{\tabcolsep}{2pt}
    \renewcommand{\arraystretch}{1.15}
    \footnotesize
	\begin{tabular}{l*4{>{\centering\arraybackslash}p{3.5cm}}}
		\toprule
		 & (1) & (2) & (3) & (4) \\
		 & F1 AI Patent & F1 AI Patent Count & F1 AI Word & F1 AI Word Count \\
		\midrule
		res & -0.021*** & -0.699*** & 0.216*** & 3.111*** \\
		& (0.006) & (0.205) & (0.011) & (0.285) \\
		\midrule
		Controls & Yes & Yes & Yes & Yes \\
		Firm FE & Yes & Yes & Yes & Yes \\
		Year FE & Yes & Yes & Yes & Yes \\
		Year\_Ind FE & Yes & Yes & Yes & Yes \\
		Year\_Prov FE & Yes & Yes & Yes & Yes \\
		$N$ & 18713 & 18713 & 18713 & 18713 \\
		Adj.\;$R^{2}$ & 0.831 & 0.869 & 0.836 & 0.857 \\
		\bottomrule
        \addlinespace[0.1em]
        \multicolumn{5}{l}{\footnotesize $^{*}$ $p<0.1$, $^{**}$ $p<0.05$, $^{***}$ $p<0.01$} \\
        \multicolumn{5}{l}{\footnotesize Notes: (1) Standard errors clustered at firm level in parentheses.} \\
	\end{tabular}\label{tabFutureValidity}
\end{table}

\autoref{tabFutureValidity} examines the predictive validity of the decoupling residual for subsequent technological and disclosure outcomes. The significantly negative coefficients on future AI patenting indicate that higher pre-policy residuals predict lower subsequent substantive innovation output. Conversely, the significantly positive coefficients on future AI narrative intensity demonstrate that higher residuals predict persistent symbolic disclosure behavior. These patterns jointly confirm that the residual measure effectively identifies firms with durable patterns of technological decoupling rather than sporadic or random deviations between disclosure and patenting.

\section{Additional Robustness Tests}\label{ApdxD}

\subsection{Subsample Excluding Year 2020}

\begin{figure}[htbp]
    \centering
    \includegraphics[width=0.65\linewidth]{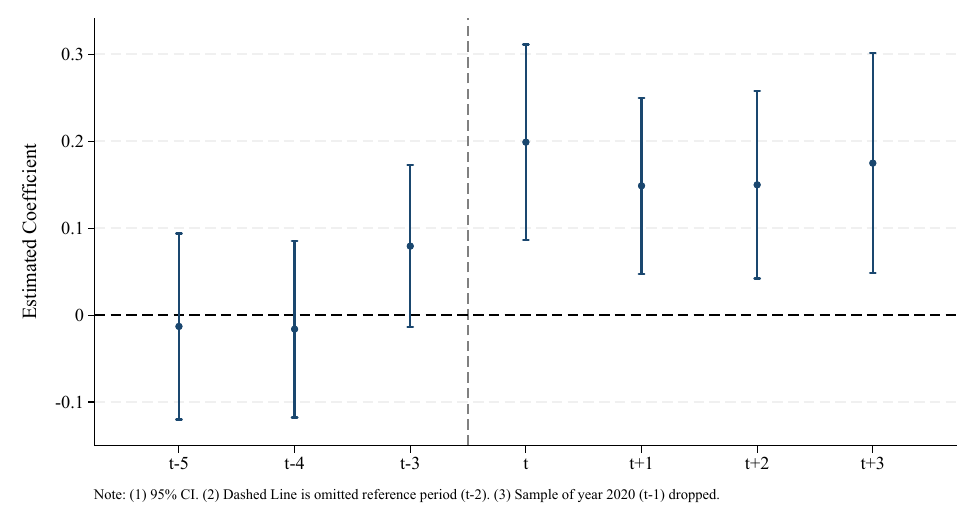}
    \caption{Parallel Trend Test (Drop Sample of 2020)}
    \label{figPTEX2020}
\end{figure}

We examine whether the treatment classification is contaminated by the COVID-19 shock or potential policy anticipation in 2020 by re-estimating the baseline using residuals averaged over 2015--2019 only. \autoref{figPTEX2020} presents the parallel trend test results after excluding 2020, which confirms the validity of the parallel trends assumption in the restricted sample.

\autoref{tabExclude2020} reports the estimation results. The coefficient increases from 0.125 to 0.156 in the full specification, suggesting that the 2020 data may have introduced measurement noise that attenuated the true effect. This confirms that the dissipation of financing advantages is driven by habitual decoupling behavior established well before the policy shock, and is not an artifact of pandemic-induced financing distortions or policy anticipation.

\begin{table}[htbp]
	\centering
	\caption{\centering Exclude the Sample from 2020}
    \setlength{\tabcolsep}{2pt}
    \renewcommand{\arraystretch}{1.15}
    \footnotesize
	\begin{tabular}{l*4{>{\centering\arraybackslash}p{3.45cm}}}
		\toprule
		 & (1) & (2) & (3) & (4) \\
		 & Debt FC & Debt FC & Debt FC & Debt FC \\
		\midrule
		AI Washing & 0.226*** & 0.177*** & 0.177*** & 0.156*** \\
		& (0.043) & (0.041) & (0.041) & (0.041) \\
        \midrule
        Controls &  & Yes & Yes & Yes \\
		Firm FE & Yes & Yes & Yes & Yes \\
		Year FE & Yes & Yes & Yes & Yes \\
		Year\_Ind FE &  &  & Yes & Yes \\
		Year\_Prov\_FE &  &  &  & Yes \\
		$N$ & 29404 & 29404 & 29404 & 29404 \\
		Adj.\;$R^{2}$ & 0.450 & 0.467 & 0.467 & 0.476 \\
		\bottomrule
        \addlinespace[0.1em]
        \multicolumn{5}{l}{\footnotesize $^{*}$ $p<0.1$, $^{**}$ $p<0.05$, $^{***}$ $p<0.01$} \\
        \multicolumn{5}{l}{\footnotesize Note: Standard errors clustered at firm level in parentheses.} \\
	\end{tabular}\label{tabExclude2020}
\end{table}

\subsection{Extreme Subsample}\label{ApdxE}

We verify in this section that the documented treatment effect is not driven by marginal observations near the classification threshold but instead reflects meaningful economic consequences for firms with severe decoupling behavior by restricting the analysis to the extreme subsample of firms with decoupling residuals in the upper tercile of the positive distribution.

\begin{table}[htbp]
	\centering
	\caption{\centering Extreme Subsample Regression}
    \setlength{\tabcolsep}{2pt}
    \renewcommand{\arraystretch}{1.15}
    \footnotesize
	\begin{tabular}{l*4{>{\centering\arraybackslash}p{3.45cm}}}
		\toprule
		 & (1) & (2) & (3) & (4) \\
		 & Debt FC & Debt FC & Debt FC & Debt FC \\
		\midrule
		AI Washing\_E & 0.260*** & 0.201*** & 0.221*** & 0.193*** \\
		& (0.052) & (0.050) & (0.050) & (0.051) \\
        \midrule
        Controls &  & Yes & Yes & Yes \\
		Firm\_FE & Yes & Yes & Yes & Yes \\
		Year\_FE & Yes & Yes & Yes & Yes \\
		Year\_Ind\_FE & & & Yes & Yes \\
		Year\_Prov\_FE & & & & Yes \\
		$N$ & 17467 & 17467 & 17461 & 17457 \\
		Adj.\;$R^{2}$ & 0.450 & 0.465 & 0.470 & 0.477 \\
		\bottomrule
        \addlinespace[0.1em]
        \multicolumn{5}{l}{\footnotesize $^{*}$ $p<0.1$, $^{**}$ $p<0.05$, $^{***}$ $p<0.01$} \\
        \multicolumn{5}{l}{\footnotesize Note: Standard errors clustered at firm level in parentheses.} \\
	\end{tabular}\label{tabExtreme}
\end{table}

\autoref{tabExtreme} reports the estimation results for this extreme subsample. The coefficient on AI washing remains positive and statistically significant across all specifications, with magnitudes ranging from 0.193 to 0.260. These estimates substantially exceed the baseline full-sample coefficient of 0.125, indicating that the financing cost penalty intensifies monotonically with the severity of pre-policy decoupling. The consistency of significance across progressively more stringent fixed effect specifications confirms that the treatment effect is robust among the most severe AI washing firms. This pattern validates that the documented dissipation of financing advantages reflects genuine market discipline applied to substantive technological decoupling rather than statistical artifacts generated by measurement error or threshold effects at the classification boundary.

\subsection{Strict Treatment Definition}\label{ApdxF}

\begin{table}[htbp]
	\centering
	\caption{\centering Strict Treatment Definition}
    \setlength{\tabcolsep}{2pt}
    \renewcommand{\arraystretch}{1.15}
    \footnotesize
	\begin{tabular}{l*4{>{\centering\arraybackslash}p{3.45cm}}}
		\toprule
		 & (1) & (2) & (3) & (4) \\
		 & Debt FC & Debt FC & Debt FC & Debt FC \\
		\midrule
		AI Washing\_S & 0.319*** & 0.131** & 0.148** & 0.134** \\
		& (0.055) & (0.058) & (0.059) & (0.062) \\
        \midrule
        Controls & & Yes & Yes & Yes \\
		Firm FE & Yes & Yes & Yes & Yes \\
		Year FE & Yes & Yes & Yes & Yes \\
		Year\_Ind\_FE &  &  & Yes & Yes \\
		Year\_Prov\_FE &  &  &  & Yes \\
		$N$ & 13487 & 13487 & 13484 & 13483 \\
		Adj.\;$R^{2}$ & 0.571 & 0.590 & 0.600 & 0.601 \\
		\bottomrule
        \addlinespace[0.1em]
        \multicolumn{5}{l}{\footnotesize $^{*}$ $p<0.1$, $^{**}$ $p<0.05$, $^{***}$ $p<0.01$} \\
        \multicolumn{5}{l}{\footnotesize Notes: (1) Strict treatment group = always positive in 2015--2020.} \\
        \multicolumn{5}{l}{\footnotesize (2) Strict control group = always negative in 2015--2020.} \\
        \multicolumn{5}{l}{\footnotesize (3) Mixed-sign firms excluded. Standard errors clustered at firm level.}
	\end{tabular}\label{tabStrict}
\end{table}

We employ a stringent definition of treatment status that requires residuals to be strictly positive in all six pre-policy years from 2015 to 2020, with the control group similarly defined as strictly negative in all years. This definition excludes firms with mixed signs, ensuring that the treatment and control groups represent firms with unambiguous and persistent decoupling behavior.

\autoref{tabStrict} reports the estimation results using the strict definition. The coefficient in the full specification is 0.134 and significant at the 5\% level, comparable to the baseline estimate of 0.125. This stability is particularly informative because the strict definition excludes 61\% of observations, retaining only firms with unambiguous persistent AI washing behavior. 

A natural concern is whether this sample attrition introduces selection bias by systematically excluding firms with transient or marginal decoupling patterns. We argue that such attrition does not threaten inference for three reasons. First, the excluded mixed-sign firms are precisely those whose decoupling status is most susceptible to measurement error. Firms with residuals oscillating around zero across years likely reflect genuine innovation volatility rather than strategic narrative manipulation, and their inclusion in the baseline mean-based definition introduces classification noise. The strict definition therefore improves measurement purity by retaining only firms whose decoupling behavior is unequivocally sustained, trading sample size for treatment clarity. Second, if the baseline estimate were driven by marginal firms with transient positive residuals, excluding them should attenuate or eliminate the effect. Instead, the coefficient remains stable and significant, indicating that the financing cost penalty is driven by firms with habitual decoupling behavior rather than sporadic outliers. Third, the direction and magnitude of the strict-definition coefficient closely track the preferred specification, suggesting that the 61\% sample reduction does not alter the underlying economic relationship but merely reduces estimation precision. We retain the mean-based definition as our preferred specification because it preserves sample representativeness and statistical power while the strict definition serves as a robustness bound confirming that results are not artifacts of ambiguous classification.

\bibliographystyle{apalike}
\bibliography{references}

@article{brown2019does,
  title={Does transparency stifle or facilitate innovation?},
  author={Brown, James R and Martinsson, Gustav},
  journal={Management Science},
  volume={65},
  number={4},
  pages={1600--1623},
  year={2019},
  publisher={INFORMS},
  doi={https://doi.org/10.1287/mnsc.2017.3002}
}

@article{xu2026illusion,
  title={The illusion of AI: how AI washing impedes green technology innovation},
  author={Xu, Zihui and Zhou, Junjun and Yuan, Guipeng and Zang, Shoujuan},
  journal={Applied Economics Letters},
  pages={1--7},
  year={2026},
  publisher={Taylor \& Francis},
  doi={https://doi.org/10.1080/13504851.2026.2648742}
}

@article{song2026ai,
  title={AI Washing: Strategic Disclosure and Backlash},
  author={Song, Xiapeng and Hou, Wenxuan and Ouyang, Zizhou and Hao, Fangmin},
  journal={Finance Research Letters},
  pages={109684},
  year={2026},
  publisher={Elsevier},
  doi={https://doi.org/10.1016/j.frl.2026.109684}
}

@article{zhou2026contractual,
  title={Contractual arrangements and information consistency: How ESG executive compensation incentives affect corporate AI disclosure},
  author={Zhou, Yiqiang and Chen, Lianghua and Zhou, Fangfang and Ye, Maoran},
  journal={Journal of Business Research},
  volume={207},
  pages={116019},
  year={2026},
  publisher={Elsevier},
  doi={https://doi.org/10.1016/j.jbusres.2026.116019}
}

@article{doh2015csr,
  title={CSR and sustainability in emerging markets: Societal, institutional, and organizational influences},
  author={Doh, Jonathan P and Littell, Benjamin and Quigley, Narda R},
  journal={Organizational Dynamics},
  volume={44},
  number={2},
  pages={112--120},
  year={2015},
  publisher={Elsevier},
  doi={http://dx.doi.org/10.1016/j.orgdyn.2015.02.005}
}

@article{kolk2016social,
  title={The social responsibility of international business: From ethics and the environment to CSR and sustainable development},
  author={Kolk, Ans},
  journal={Journal of World Business},
  volume={51},
  number={1},
  pages={23--34},
  year={2016},
  publisher={Elsevier},
  doi={https://doi.org/10.1016/j.jwb.2015.08.010}
}

@article{kumar2021cross,
  title={Cross-national differences in stakeholder management: Applying institutional theory and comparative capitalism framework},
  author={Kumar, Kamalesh and Boesso, Giacomo and Batra, Rishtee and Yao, Jun},
  journal={Business Strategy and the Environment},
  volume={30},
  number={5},
  pages={2354--2366},
  year={2021},
  publisher={Wiley Online Library},
  doi={https://doi.org/10.1002/bse.2750}
}

@article{rozeboom2025corporate,
  title={Corporate Sincerity: Accommodation, Reputation Washing, and Moral Credit: GJ Rozeboom},
  author={Rozeboom, Grant J},
  journal={Journal of Business Ethics},
  volume={202},
  number={3},
  pages={503--517},
  year={2025},
  publisher={Springer},
  doi={https://doi.org/10.1007/s10551-025-06000-1}
}

@article{pope2016csr,
  title={CSR-washing is rare: A conceptual framework, literature review, and critique},
  author={Pope, Shawn and W{\ae}raas, Arild},
  journal={Journal of Business Ethics},
  volume={137},
  number={1},
  pages={173--193},
  year={2016},
  publisher={Springer},
  doi={https://doi.org/10.1007/s10551-015-2546-z}
}

@article{crilly2012faking,
  title={Faking it or muddling through? Understanding decoupling in response to stakeholder pressures},
  author={Crilly, Donal and Zollo, Maurizio and Hansen, Morten T},
  journal={Academy of Management Journal},
  volume={55},
  number={6},
  pages={1429--1448},
  year={2012},
  publisher={American Society of Nephrology},
  doi={https://doi.org/10.5465/amj.2010.0697}
}

@article{zahra1993environment,
  title={Environment, corporate entrepreneurship, and financial performance: A taxonomic approach},
  author={Zahra, Shaker A},
  journal={Journal of Business Venturing},
  volume={8},
  number={4},
  pages={319--340},
  year={1993},
  publisher={Elsevier},
  doi={https://doi.org/10.1016/0883-9026(93)90003-N}
}

@article{scherer2020corporate,
  title={Corporate governance for responsible innovation: Approaches to corporate governance and their implications for sustainable development},
  author={Scherer, Andreas Georg and Voegtlin, Christian},
  journal={Academy of Management Perspectives},
  volume={34},
  number={2},
  pages={182--208},
  year={2020},
  publisher={Academy of Management Briarcliff Manor},
  doi={https://doi.org/10.5465/amp.2017.0175}
}

@article{li2022can,
  title={How can China's sustainable development be damaged in consequence of financial misallocation? Analysis from the perspective of regional innovation capability},
  author={Li, Yunwei and Long, Wenjing and Ning, Xiao and Zhu, Yumeng and Guo, Yifan and Huang, Zhou and Hao, Yu},
  journal={Business Strategy and the Environment},
  volume={31},
  number={7},
  pages={3649--3668},
  year={2022},
  publisher={Wiley Online Library},
  doi={https://doi.org/10.1002/bse.3113}
}

@article{liu2026impact,
  title={The Impact of AI Washing on Enterprises’ Access to Bank Loans: From the Perspective of External Governance},
  author={Liu, Weiqi and Li, Meifang},
  journal={Finance Research Letters},
  pages={109884},
  year={2026},
  publisher={Elsevier},
  doi={https://doi.org/10.1016/j.frl.2026.109884}
}

@article{sun2026unveiling,
  title={Unveiling AI washing: Bridging corporate technological gaps through a cognitive dissonance lens},
  author={Sun, Zhe and Wen, Yujun and Zhao, Liang and Almugren, Intesar and Galgotia, Aradhana},
  journal={Technological Forecasting and Social Change},
  volume={225},
  pages={124511},
  year={2026},
  publisher={Elsevier},
  doi={https://doi.org/10.1016/j.techfore.2025.124511}
}

@article{gong2021punishment,
  title={Punishment by securities regulators, corporate social responsibility and the cost of debt},
  author={Gong, Guangming and Huang, Xin and Wu, Sirui and Tian, Haowen and Li, Wanjin},
  journal={Journal of Business Ethics},
  volume={171},
  number={2},
  pages={337--356},
  year={2021},
  publisher={Springer},
  doi={https://doi.org/10.1007/s10551-020-04438-z}
}

@article{liu2022capital,
  title={The capital market responses to new energy vehicle (NEV) subsidies: An event study on China},
  author={Liu, Chang and Liu, Yuan and Zhang, Dayong and Xie, Chunping},
  journal={Energy Economics},
  volume={105},
  pages={105677},
  year={2022},
  publisher={Elsevier},
  doi={https://doi.org/10.1016/j.eneco.2021.105677}
}

@article{marquis2016scrutiny,
  title={Scrutiny, norms, and selective disclosure: A global study of greenwashing},
  author={Marquis, Christopher and Toffel, Michael W and Zhou, Yanhua},
  journal={Organization Science},
  volume={27},
  number={2},
  pages={483--504},
  year={2016},
  publisher={Informs},
  doi={https://doi.org/10.1287/orsc.2015.1039}
}

@article{takalo2010adverse,
  title={Adverse selection and financing of innovation: is there a need for R\&D subsidies?},
  author={Takalo, Tuomas and Tanayama, Tanja},
  journal={The Journal of Technology Transfer},
  volume={35},
  number={1},
  pages={16--41},
  year={2010},
  publisher={Springer},
  doi={https://doi.org/10.1007/s10961-009-9112-8}
}

@article{keister2004capital,
  title={Capital structure in transition: The transformation of financial strategies in China's emerging economy},
  author={Keister, Lisa A},
  journal={Organization Science},
  volume={15},
  number={2},
  pages={145--158},
  year={2004},
  publisher={INFORMS},
  doi={https://doi.org/10.1287/orsc.1040.0043}
}

@article{cong2019credit,
  title={Credit allocation under economic stimulus: Evidence from China},
  author={Cong, Lin William and Gao, Haoyu and Ponticelli, Jacopo and Yang, Xiaoguang},
  journal={The Review of Financial Studies},
  volume={32},
  number={9},
  pages={3412--3460},
  year={2019},
  publisher={Oxford University Press},
  doi={https://doi.org/10.1093/rfs/hhz008}
}

@article{wang2022public,
  title={Public attention and investment efficiency: Incentive effect or deterrent effect? Analysis on heterogeneous bilateral stochastic frontier model},
  author={Wang, Jingjuan and Xia, Weili},
  journal={Technological Forecasting and Social Change},
  volume={185},
  pages={122043},
  year={2022},
  publisher={Elsevier},
  doi={https://doi.org/10.1016/j.techfore.2022.122043}
}

@article{stern2023china,
  title={China’s new growth story: Linking the 14th Five-Year Plan with the 2060 carbon neutrality pledge},
  author={Stern, Nicholas and Xie, Chunping},
  journal={Journal of Chinese Economic and Business Studies},
  volume={21},
  number={1},
  pages={5--25},
  year={2023},
  publisher={Taylor \& Francis},
  doi={https://doi.org/10.1080/14765284.2022.2073172}
}

@article{merkley2014narrative,
  title={Narrative disclosure and earnings performance: Evidence from R\&D disclosures},
  author={Merkley, Kenneth J},
  journal={The Accounting Review},
  volume={89},
  number={2},
  pages={725--757},
  year={2014},
  publisher={American Accounting Association},
  doi={https://doi.org/10.2308/accr-50649}
}

@article{bester1985screening,
  title={Screening vs. rationing in credit markets with imperfect information},
  author={Bester, Helmut},
  journal={The American Economic Review},
  volume={75},
  number={4},
  pages={850--855},
  year={1985},
  publisher={JSTOR},
  doi={https://www.jstor.org/stable/1821362}
}

@article{demerjian2024positive,
  title={A positive theory of information for debt contracting: Implications for financial reporting},
  author={Demerjian, Peter},
  journal={Journal of Business Finance \& Accounting},
  year={2024},
  publisher={Wiley Online Library},
  doi={https://doi.org/10.1111/jbfa.70038}
}

@article{matolcsy2008association,
  title={The association between technological conditions and the market value of equity},
  author={Matolcsy, Zoltan P and Wyatt, Anne},
  journal={The Accounting Review},
  volume={83},
  number={2},
  pages={479--518},
  year={2008},
  doi={https://doi.org/10.2308/accr.2008.83.2.479}
}

@article{wei2022r,
  title={R\&D investment and debt financing of high-tech firms in emerging economies: The role of patents and state ownership},
  author={Wei, Haixiao and Xie, En and Gao, Jingzhe},
  journal={IEEE Transactions on Engineering Management},
  volume={71},
  pages={753--770},
  year={2022},
  publisher={IEEE},
  doi={https://doi.org/10.1109/TEM.2021.3133330}
}

@article{bernini2025measuring,
  title={Measuring machinewashing under the corporate digital responsibility theory: A proposal for a methodological path},
  author={Bernini, Francesca and Ferretti, Paola and Gonnella, Cristina and La Rosa, Fabio},
  journal={Business Ethics, the Environment \& Responsibility},
  volume={34},
  number={2},
  pages={328--346},
  year={2025},
  publisher={Wiley Online Library},
  doi={https://doi.org/10.1111/beer.12653}
}

@article{seele2022greenwashing,
  title={From greenwashing to machinewashing: A model and future directions derived from reasoning by analogy},
  author={Seele, Peter and Schultz, Mario D},
  journal={Journal of Business Ethics},
  volume={178},
  number={4},
  pages={1063--1089},
  year={2022},
  publisher={Springer},
  doi={https://doi.org/10.1007/s10551-022-05054-9}
}

@article{meyer1977institutionalized,
  title={Institutionalized organizations: Formal structure as myth and ceremony},
  author={Meyer, John W and Rowan, Brian},
  journal={American Journal of Sociology},
  volume={83},
  number={2},
  pages={340--363},
  year={1977},
  publisher={University of Chicago Press},
  doi={https://doi.org/10.1086/226550}
}

@article{behnam2011accountability,
  title={Where is the accountability in international accountability standards?: A decoupling perspective},
  author={Behnam, Michael and MacLean, Tammy L},
  journal={Business Ethics Quarterly},
  volume={21},
  number={1},
  pages={45--72},
  year={2011},
  publisher={Cambridge University Press},
  doi={https://doi.org/10.5840/beq20112113}
}

@article{shi2018regulatory,
  title={Is regulatory adoption ceremonial? E vidence from lead director appointments},
  author={Shi, Wei and Connelly, Brian L},
  journal={Strategic Management Journal},
  volume={39},
  number={8},
  pages={2386--2413},
  year={2018},
  publisher={Wiley Online Library},
  doi={https://doi.org/10.1002/smj.2901}
}

@article{testa2018internalization,
  title={Internalization of environmental practices and institutional complexity: can stakeholders pressures encourage greenwashing?},
  author={Testa, Francesco and Boiral, Olivier and Iraldo, Fabio},
  journal={Journal of Business Ethics},
  volume={147},
  number={2},
  pages={287--307},
  year={2018},
  publisher={Springer},
  doi={https://doi.org/10.1007/s10551-015-2960-2}
}

@article{xing2026ai,
  title={AI technology, AI narrative, and firm value},
  author={Xing, Xiaoqiang and Zhang, Zhu and He, Weixuan},
  journal={Technovation},
  volume={149},
  pages={103349},
  year={2026},
  publisher={Elsevier},
  doi={https://doi.org/10.1016/j.technovation.2025.103349}
}

@article{katila2008effects,
  title={Effects of search timing on innovation: The value of not being in sync with rivals},
  author={Katila, Riitta and Chen, Eric L},
  journal={Administrative Science Quarterly},
  volume={53},
  number={4},
  pages={593--625},
  year={2008},
  publisher={SAGE Publications Sage CA},
  doi={https://doi.org/10.2189/asqu.53.4.593}
}

@article{garcia2005uses,
  title={Uses of agent-based modeling in innovation/new product development research},
  author={Garcia, Rosanna},
  journal={Journal of Product Innovation Management},
  volume={22},
  number={5},
  pages={380--398},
  year={2005},
  publisher={Wiley Online Library},
  doi={https://doi.org/10.1111/j.1540-5885.2005.00136.x}
}

@article{payne2006examining,
  title={Examining configurations and firm performance in a suboptimal equifinality context},
  author={Payne, G Tyge},
  journal={Organization Science},
  volume={17},
  number={6},
  pages={756--770},
  year={2006},
  publisher={INFORMS},
  doi={https://doi.org/10.1287/orsc.1060.0203}
}

@article{busenbark2017foreshadowing,
  title={Foreshadowing as impression management: Illuminating the path for security analysts},
  author={Busenbark, John R and Lange, Donald and Certo, S Trevis},
  journal={Strategic Management Journal},
  volume={38},
  number={12},
  pages={2486--2507},
  year={2017},
  publisher={Wiley Online Library},
  doi={https://doi.org/10.1002/smj.2659}
}

@article{argyres2004r,
  title={R\&D, organization structure, and the development of corporate technological knowledge},
  author={Argyres, Nicholas S and Silverman, Brian S},
  journal={Strategic Management Journal},
  volume={25},
  number={8-9},
  pages={929--958},
  year={2004},
  publisher={Wiley Online Library},
  doi={https://doi.org/10.1002/smj.387}
}

@article{argyres2020organizational,
  title={Organizational change and the dynamics of innovation: Formal R\&D structure and intrafirm inventor networks},
  author={Argyres, Nicholas and Rios, Luis A and Silverman, Brian S},
  journal={Strategic Management Journal},
  volume={41},
  number={11},
  pages={2015--2049},
  year={2020},
  publisher={Wiley Online Library},
  doi={https://doi.org/10.1002/smj.3217}
}

@article{sartori2023minding,
  title={Minding the gap(s): public perceptions of AI and socio-technical imaginaries},
  author={Sartori, Laura and Bocca, Giulia},
  journal={AI \& Society},
  volume={38},
  number={2},
  pages={443--458},
  year={2023},
  publisher={Springer},
  doi={https://doi.org/10.1007/s00146-022-01422-1}
}

@article{lajoie2025content,
  title={Content marketing as a propaganda vehicle for a romantic-managerial conception of artificial intelligence},
  author={Lajoie, Pier-Luc},
  journal={Critical Perspectives on Accounting},
  volume={102},
  pages={102810},
  year={2025},
  publisher={Elsevier},
  doi={https://doi.org/10.1016/j.cpa.2025.102810}
}

@article{carey2026regulating,
  title={Regulating uncertainty: Governing general-purpose ai models and systemic risk},
  author={Carey, Samuel},
  journal={European Journal of Risk Regulation},
  volume={17},
  number={1},
  pages={123--139},
  year={2026},
  publisher={Cambridge University Press},
  doi={https://doi.org/10.1017/err.2025.10040}
}

@article{schmitt2022mapping,
  title={Mapping global AI governance: a nascent regime in a fragmented landscape},
  author={Schmitt, Lewin},
  journal={AI and Ethics},
  volume={2},
  number={2},
  pages={303--314},
  year={2022},
  publisher={Springer},
  doi={https://doi.org/10.1007/s43681-021-00083-y}
}

@article{yu2020greenwashing,
  title={Greenwashing in environmental, social and governance disclosures},
  author={Yu, Ellen Pei-yi and Van Luu, Bac and Chen, Catherine Huirong},
  journal={Research in International Business and Finance},
  volume={52},
  pages={101192},
  year={2020},
  publisher={Elsevier},
  doi={https://doi.org/10.1016/j.ribaf.2020.101192}
}

@article{hirshleifer2003limited,
  title={Limited attention, information disclosure, and financial reporting},
  author={Hirshleifer, David and Teoh, Siew Hong},
  journal={Journal of Accounting and Economics},
  volume={36},
  number={1-3},
  pages={337--386},
  year={2003},
  publisher={Elsevier},
  doi={https://doi.org/10.1016/j.jacceco.2003.10.002}
}

@article{loewenstein2025economics,
  title={The economics of attention},
  author={Loewenstein, George and Wojtowicz, Zachary},
  journal={Journal of Economic Literature},
  volume={63},
  number={3},
  pages={1038--1089},
  year={2025},
  publisher={American Economic Association},
  doi={https://doi.org/10.1257/jel.20241665}
}

@article{mackowiak2023rational,
  title={Rational inattention: A review},
  author={Ma{\'c}kowiak, Bartosz and Mat{\v{e}}jka, Filip and Wiederholt, Mirko},
  journal={Journal of Economic Literature},
  volume={61},
  number={1},
  pages={226--273},
  year={2023},
  publisher={American Economic Association},
  doi={https://doi.org/10.1257/jel.20211524}
  
}

@article{huynh2023panic,
  title={Panic selling when disaster strikes: Evidence in the bond and stock markets},
  author={Huynh, Thanh D and Xia, Ying},
  journal={Management Science},
  volume={69},
  number={12},
  pages={7448--7467},
  year={2023},
  publisher={INFORMS},
  doi={https://doi.org/10.1287/mnsc.2021.4018}
}

@article{bertelli2015mass,
  title={Mass administrative reorganization, media attention, and the paradox of information},
  author={Bertelli, Anthony M and Sinclair, J Andrew},
  journal={Public Administration Review},
  volume={75},
  number={6},
  pages={855--866},
  year={2015},
  publisher={Wiley Online Library},
  doi={https://doi.org/10.1111/puar.12396}
}

@article{abad2019informational,
  title={Informational role of rating revisions after reputational events and regulation reforms},
  author={Abad, Pilar and Ferreras, Rodrigo and Robles, M-Dolores},
  journal={International Review of Financial Analysis},
  volume={62},
  pages={91--103},
  year={2019},
  publisher={Elsevier},
  doi={https://doi.org/10.1016/j.irfa.2019.01.005}
}

@article{wirtz2020dark,
  title={The dark sides of artificial intelligence: An integrated AI governance framework for public administration},
  author={Wirtz, Bernd W and Weyerer, Jan C and Sturm, Benjamin J},
  journal={International Journal of Public Administration},
  volume={43},
  number={9},
  pages={818--829},
  year={2020},
  publisher={Taylor \& Francis},
  doi={https://doi.org/10.1080/01900692.2020.1749851}
}

@article{af2023discursive,
  title={Discursive framing and organizational venues: mechanisms of artificial intelligence policy adoption},
  author={Af Malmborg, Frans and Trondal, Jarle},
  journal={International Review of Administrative Sciences},
  volume={89},
  number={1},
  pages={39--58},
  year={2023},
  publisher={Sage Publications Sage UK},
  doi={https://doi.org/10.1177/00208523211007533}
}

@article{dunleavy2025data,
  title={Data science, artificial intelligence and the third wave of digital era governance},
  author={Dunleavy, Patrick and Margetts, Helen},
  journal={Public Policy and Administration},
  volume={40},
  number={2},
  pages={185--214},
  year={2025},
  publisher={SAGE Publications Sage UK},
  doi={https://doi.org/10.1177/09520767231198737}
}

@article{federico2025trade,
  title={Trade shocks and credit reallocation},
  author={Federico, Stefano and Hassan, Fadi and Rappoport, Veronica},
  journal={The American Economic Review},
  volume={115},
  number={4},
  pages={1142--1169},
  year={2025},
  doi={https://doi.org/10.1257/aer.20200704}
}

@article{stiglitz1981credit,
  title={Credit rationing in markets with imperfect information},
  author={Stiglitz, Joseph E and Weiss, Andrew},
  journal={The American Economic Review},
  volume={71},
  number={3},
  pages={393--410},
  year={1981},
  publisher={JSTOR},
  url={https://www.jstor.org/stable/1802787}
}

@article{healy2001information,
  title={Information asymmetry, corporate disclosure, and the capital markets: A review of the empirical disclosure literature},
  author={Healy, Paul M and Palepu, Krishna G},
  journal={Journal of Accounting and Economics},
  volume={31},
  number={1-3},
  pages={405--440},
  year={2001},
  publisher={Elsevier},
  doi={https://doi.org/10.1016/S0165-4101(01)00018-0}
}

@article{moss2015effect,
  title={The effect of virtuous and entrepreneurial orientations on microfinance lending and repayment: A signaling theory perspective},
  author={Moss, Todd W and Neubaum, Donald O and Meyskens, Moriah},
  journal={Entrepreneurship Theory and Practice},
  volume={39},
  number={1},
  pages={27--52},
  year={2015},
  publisher={SAGE Publications Sage CA},
  doi={https://doi.org/10.1111/etap.12110}
}

@article{mancusi2014r,
  title={R\&D and credit rationing in SMEs},
  author={Mancusi, Maria Luisa and Vezzulli, Andrea},
  journal={Economic Inquiry},
  volume={52},
  number={3},
  pages={1153--1172},
  year={2014},
  publisher={Wiley Online Library},
  doi={https://doi.org/10.1111/ecin.12080}
}

@article{cerqueiro2017debtor,
  title={Debtor rights, credit supply, and innovation},
  author={Cerqueiro, Geraldo and Hegde, Deepak and Penas, Mar{\'\i}a Fabiana and Seamans, Robert C},
  journal={Management Science},
  volume={63},
  number={10},
  pages={3311--3327},
  year={2017},
  publisher={INFORMS},
  doi={https://doi.org/10.1287/mnsc.2016.2509}
}

@article{krishnan2009recent,
  title={Recent trends in audit report and earnings announcement lags},
  author={Krishnan, Jayanthi and Yang, Joon S},
  journal={Accounting Horizons},
  volume={23},
  number={3},
  pages={265--288},
  year={2009},
  publisher={American Accounting Association},
  doi={https://doi.org/10.2308/acch.2009.23.3.265}
}

@article{hottenrott2016patents,
  title={Patents as quality signals? The implications for financing constraints on R\&D},
  author={Hottenrott, Hanna and Hall, Bronwyn H and Czarnitzki, Dirk},
  journal={Economics of Innovation and New Technology},
  volume={25},
  number={3},
  pages={197--217},
  year={2016},
  publisher={Taylor \& Francis},
  doi={https://doi.org/10.1080/10438599.2015.1076200}
}

@article{hong2017reading,
  title={Reading the 13th five-year plan: Reflections on China’s ICT policy},
  author={Hong, Yu},
  journal={International Journal of Communication},
  volume={11},
  pages={24--24},
  year={2017},
  doi={https://ssrn.com/abstract=2953044}
}

@article{roberts2021chinese,
  title={The Chinese approach to artificial intelligence: an analysis of policy, ethics, and regulation},
  author={Roberts, Huw and Cowls, Josh and Morley, Jessica and Taddeo, Mariarosaria and Wang, Vincent and Floridi, Luciano},
  journal={AI \& Society},
  volume={36},
  number={1},
  pages={59--77},
  year={2021},
  publisher={Springer},
  doi={https://doi.org/10.1007/s00146-020-00992-2}
}

@article{khanal2025development,
  title={Development of new generation of artificial intelligence in China: When Beijing’s global ambitions meet local realities},
  author={Khanal, Shaleen and Zhang, Hongzhou and Taeihagh, Araz},
  journal={Journal of Contemporary China},
  volume={34},
  number={151},
  pages={19--42},
  year={2025},
  publisher={Taylor \& Francis},
  doi={https://doi.org/10.1080/10670564.2024.2333492}
}

@article{poo2021innovation,
  title={Innovation and reform: China's 14th five-year plan unfolds},
  author={Poo, Mu-ming},
  journal={National Science Review},
  volume={8},
  number={1},
  pages={nwaa294},
  year={2021},
  publisher={Oxford University Press},
  doi={https://doi.org/10.1093/nsr/nwaa294}
}

@article{JENSEN1976305,
    title = {Theory of the firm: Managerial behavior, agency costs and ownership structure},
    author = {Michael C. Jensen and William H. Meckling},
    journal = {Journal of Financial Economics},
    volume = {3},
    number = {4},
    pages = {305-360},
    year = {1976},
    doi={https://doi.org/10.1016/0304-405X(76)90026-X}
}

@article{xin1996guanxi,
  title={Guanxi: Connections as substitutes for formal institutional support},
  author={Xin, Katherine K and Pearce, Jone L},
  journal={Academy of Management Journal},
  volume={39},
  number={6},
  pages={1641--1658},
  year={1996},
  publisher={Academy of Management Briarcliff Manor},
  doi={https://doi.org/10.5465/257072}
}

@article{bharath2008accounting,
  title={Accounting quality and debt contracting},
  author={Bharath, Sreedhar T and Sunder, Jayanthi and Sunder, Shyam V},
  journal={The Accounting Review},
  volume={83},
  number={1},
  pages={1--28},
  year={2008},
  publisher={American Accounting Association},
  doi={https://doi.org/10.2308/accr.2008.83.1.1}
}

@article{balakrishnan2019bank,
  title={Bank asset transparency and credit supply},
  author={Balakrishnan, Karthik and Ertan, Aytekin},
  journal={The Review of Accounting Studies},
  volume={24},
  number={4},
  pages={1359--1391},
  year={2019},
  publisher={Springer},
  doi={https://doi.org/10.1007/s11142-019-09510-2}
}

@article{zhou2003embeddedness,
  title={Embeddedness and contractual relationships in China's transitional economy},
  author={Zhou, Xueguang and Zhao, Wei and Li, Qiang and Cai, He},
  journal={American Sociological Review},
  volume={68},
  number={1},
  pages={75--102},
  year={2003},
  publisher={Sage Publications Sage CA},
  doi={https://doi.org/10.1177/000312240306800104}
}

@article{wang2022government,
  title={Government intervention and debt financing costs: Evidence from government-guided funds},
  author={Wang, Kegui and Wang, Juxian and Dong, Nan},
  journal={Emerging Markets Finance And Trade},
  volume={58},
  number={11},
  pages={3284--3296},
  year={2022},
  publisher={Taylor \& Francis},
  doi={https://doi.org/10.1080/1540496X.2022.2042249}
}

@article{zou2008debt,
  title={Debt capacity, cost of debt, and corporate insurance},
  author={Zou, Hong and Adams, Mike B},
  journal={Journal of Financial and Quantitative Analysis},
  volume={43},
  number={2},
  pages={433--466},
  year={2008},
  publisher={Cambridge University Press},
  doi={https://doi.org/10.1017/S0022109000003586}
}

@article{athey2017state,
  title={The state of applied econometrics: Causality and policy evaluation},
  author={Athey, Susan and Imbens, Guido W},
  journal={Journal of Economic Perspectives},
  volume={31},
  number={2},
  pages={3--32},
  year={2017},
  publisher={American Economic Association},
  doi={https://doi.org/10.1257/jep.31.2.3}
}

@article{li2025impact,
  title={Impact of artificial intelligence on corporate green transformation},
  author={Li, Huanyu and Wu, Hao and Rao, Jian},
  journal={Finance Research Letters},
  volume={80},
  pages={107427},
  year={2025},
  publisher={Elsevier},
  doi={https://doi.org/10.1016/j.frl.2025.107427}
}

@article{chen2009does,
  title={Does the type of ownership control matter? Evidence from China’s listed companies},
  author={Chen, Gongmeng and Firth, Michael and Xu, Liping},
  journal={Journal of Banking \& Finance},
  volume={33},
  number={1},
  pages={171--181},
  year={2009},
  publisher={Elsevier},
  doi={https://doi.org/10.1016/j.jbankfin.2007.12.023}
}

@article{coles2023empirical,
  title={An empirical assessment of empirical corporate finance},
  author={Coles, Jeffrey L and Li, Zhichuan F},
  journal={Journal of Financial and Quantitative Analysis},
  volume={58},
  number={4},
  pages={1391--1430},
  year={2023},
  publisher={Cambridge University Press},
  doi={https://doi.org/10.1017/S0022109022000448}
}

@article{mitton2022methodological,
  title={Methodological variation in empirical corporate finance},
  author={Mitton, Todd},
  journal={The Review of Financial Studies},
  volume={35},
  number={2},
  pages={527--575},
  year={2022},
  publisher={Oxford University Press},
  doi={https://doi.org/10.1093/rfs/hhab030}
}

@article{jin2023real,
  title={The real effects of implicit government guarantee: evidence from Chinese state-owned enterprise defaults},
  author={Jin, Shuang and Wang, Wei and Zhang, Zilong},
  journal={Management Science},
  volume={69},
  number={6},
  pages={3650--3674},
  year={2023},
  publisher={INFORMS},
  doi={https://doi.org/10.1287/mnsc.2022.4483}
}

@article{yang2025employee,
  title={Employee overtime and innovation dilemma},
  author={Yang, Jingjing and Fan, Di and Li, Caifu},
  journal={Journal of Business Ethics},
  volume={200},
  number={3},
  pages={689--713},
  year={2025},
  publisher={Springer},
  doi={https://doi.org/10.1007/s10551-024-05918-2}
}

@article{lai2023judicial,
  title={Judicial independence and corporate innovation: Evidence from the establishment of circuit courts},
  author={Lai, Shaojie and Yang, Laifeng and Wang, Qing and Anderson, Hamish D},
  journal={Journal of Corporate Finance},
  volume={80},
  pages={102424},
  year={2023},
  publisher={Elsevier},
  doi={https://doi.org/10.1016/j.jcorpfin.2023.102424}
}

@article{gunny2010relation,
  title={The relation between earnings management using real activities manipulation and future performance: Evidence from meeting earnings benchmarks},
  author={Gunny, Katherine A},
  journal={Contemporary Accounting Research},
  volume={27},
  number={3},
  pages={855--888},
  year={2010},
  publisher={Wiley Online Library},
  doi={https://doi.org/10.1111/j.1911-3846.2010.01029.x}
}

@article{bereskin2018real,
  title={The real effects of real earnings management: Evidence from innovation},
  author={Bereskin, Frederick L and Hsu, Po-Hsuan and Rotenberg, Wendy},
  journal={Contemporary Accounting Research},
  volume={35},
  number={1},
  pages={525--557},
  year={2018},
  publisher={Wiley Online Library},
  doi={https://doi.org/10.1111/1911-3846.12376}
}

@article{vorst2016real,
  title={Real earnings management and long-term operating performance: The role of reversals in discretionary investment cuts},
  author={Vorst, Patrick},
  journal={The Accounting Review},
  volume={91},
  number={4},
  pages={1219--1256},
  year={2016},
  publisher={American Accounting Association},
  doi={https://doi.org/10.2308/accr-51281}
}

@article{rosenbaum1983central,
  title={The central role of the propensity score in observational studies for causal effects},
  author={Rosenbaum, Paul R and Rubin, Donald B},
  journal={Biometrika},
  volume={70},
  number={1},
  pages={41--55},
  year={1983},
  publisher={Oxford University Press},
  doi={https://doi.org/10.1093/biomet/70.1.41}
}

@article{hainmueller2012entropy,
  title={Entropy balancing for causal effects: A multivariate reweighting method to produce balanced samples in observational studies},
  author={Hainmueller, Jens},
  journal={Political Analysis},
  volume={20},
  number={1},
  pages={25--46},
  year={2012},
  publisher={Cambridge University Press},
  doi={https://doi.org/10.1093/pan/mpr025}
}

@article{abadie2010synthetic,
  title={Synthetic control methods for comparative case studies: Estimating the effect of California’s tobacco control program},
  author={Abadie, Alberto and Diamond, Alexis and Hainmueller, Jens},
  journal={Journal of the American Statistical Association},
  volume={105},
  number={490},
  pages={493--505},
  year={2010},
  publisher={Taylor \& Francis},
  doi={https://doi.org/10.1198/jasa.2009.ap08746}
}

@article{jiang2024does,
  title={Does knowledge of digital technology affect corporate innovation? Evidence from CEOs with digital technology backgrounds in China},
  author={Jiang, Anxuan and Ma, Jianteng and Wang, Zhitao and Zhou, Ming},
  journal={Applied Economics},
  volume={56},
  number={49},
  pages={5939--5956},
  year={2024},
  publisher={Taylor \& Francis},
  doi={https://doi.org/10.1080/00036846.2023.2266605}
}

@article{gore2011role,
  title={The role of technical expertise in firm governance structure: Evidence from chief financial officer contractual incentives},
  author={Gore, Angela K and Matsunaga, Steve and Eric Yeung, P},
  journal={Strategic Management Journal},
  volume={32},
  number={7},
  pages={771--786},
  year={2011},
  publisher={Wiley Online Library},
  doi={https://doi.org/10.1002/smj.907}
}

@article{czarnitzki2009capital,
  title={Capital control, debt financing and innovative activity},
  author={Czarnitzki, Dirk and Kraft, Kornelius},
  journal={Journal of Economic Behavior \& Organization},
  volume={71},
  number={2},
  pages={372--383},
  year={2009},
  publisher={Elsevier},
  doi={https://doi.org/10.1016/j.jebo.2009.03.017}
}

@article{tan2025go,
  title={Go green: How does Green Credit Policy promote corporate green transformation in China},
  author={Tan, Weijie and Yan, Edward Hengzhou and Yip, Wai Sze},
  journal={Journal of International Financial Management \& Accounting},
  volume={36},
  number={1},
  pages={38--67},
  year={2025},
  publisher={Wiley Online Library},
  doi={https://doi.org/10.1111%2Fjifm.12218}
}

@article{baxter1967leverage,
  title={Leverage, risk of ruin and the cost of capital},
  author={Baxter, Nevins D},
  journal={The Journal of Finance},
  volume={22},
  number={3},
  pages={395--403},
  year={1967},
  publisher={JSTOR},
  doi={https://doi.org/10.2307/2978892}
}

@article{jiang2008beating,
  title={Beating earnings benchmarks and the cost of debt},
  author={Jiang, John},
  journal={The Accounting Review},
  volume={83},
  number={2},
  pages={377--416},
  year={2008},
  doi = {https://doi.org/10.2308/accr.2008.83.2.377}
}

@article{vanacker2010pecking,
  title={Pecking order and debt capacity considerations for high-growth companies seeking financing},
  author={Vanacker, Tom R and Manigart, Sophie},
  journal={Small Business Economics},
  volume={35},
  number={1},
  pages={53--69},
  year={2010},
  publisher={Springer},
  doi={https://doi.org/10.1007/s11187-008-9150-x}
}

@article{sanchez2011ownership,
  title={Ownership structure and the cost of debt},
  author={S{\'a}nchez-Ballesta, Juan Pedro and Garc{\'\i}a-Meca, Emma},
  journal={European Accounting Review},
  volume={20},
  number={2},
  pages={389--416},
  year={2011},
  publisher={Taylor \& Francis},
  doi={https://doi.org/10.1080/09638180903487834}
}

@article{zellner1962efficient,
  title={An efficient method of estimating seemingly unrelated regressions and tests for aggregation bias},
  author={Zellner, Arnold},
  journal={Journal of the American statistical Association},
  volume={57},
  number={298},
  pages={348--368},
  year={1962},
  publisher={Taylor \& Francis},
  doi={https://doi.org/10.1080/01621459.1962.10480664}
}

@article{xie2025does,
  title={Does vertical supervision enhance labor productivity? Evidence from China’s central environmental protection inspection},
  author={Xie, Weimin and Guo, Jialu and Zhang, Hengxin and Fang, Mingxiao},
  journal={Humanities and Social Sciences Communications},
  volume={12},
  number={1},
  pages={1--15},
  year={2025},
  publisher={Palgrave Macmillan},
  doi={https://doi.org/10.1057/s41599-025-04958-x}
}

@article{agarwal2010distance,
  title={Distance and private information in lending},
  author={Agarwal, Sumit and Hauswald, Robert},
  journal={The Review of Financial Studies},
  volume={23},
  number={7},
  pages={2757--2788},
  year={2010},
  publisher={Oxford University Press},
  doi={https://doi.org/10.1093/rfs/hhq001}
}

@article{del2020soft,
  title={Soft information production in SME lending},
  author={Del Gaudio, Belinda L and Griffiths, Mark D and Sampagnaro, Gabriele},
  journal={Journal of Financial Research},
  volume={43},
  number={1},
  pages={121--151},
  year={2020},
  publisher={Wiley Online Library},
  doi={https://doi.org/10.1111/jfir.12198}
}

@article{karlan2009observing,
  title={Observing unobservables: Identifying information asymmetries with a consumer credit field experiment},
  author={Karlan, Dean and Zinman, Jonathan},
  journal={Econometrica},
  volume={77},
  number={6},
  pages={1993--2008},
  year={2009},
  publisher={Wiley Online Library}
}

@article{hong2021lender,
  title={Lender monitoring and the efficacy of managerial risk-taking incentives},
  author={Hong, Hyun A and Ryou, Ji Woo and Srivastava, Anup},
  journal={The Accounting Review},
  volume={96},
  number={4},
  pages={315--339},
  year={2021},
  publisher={American Accounting Association}
}

@article{cheng2008analyst,
  title={Analyst following and credit ratings},
  author={Cheng, Mei and Subramanyam, KR},
  journal={Contemporary Accounting Research},
  volume={25},
  number={4},
  pages={1007--1044},
  year={2008},
  publisher={Wiley Online Library}
}

\end{document}